\documentclass[twocolumn,preprintnumbers,amsmath,amssymb,superscriptaddress,nofootinbib,longbibliography,prl]{revtex4-2}  

\usepackage[utf8]{inputenc}
\usepackage[english]{babel}

\usepackage{graphicx}

\usepackage{leftidx}  
\usepackage{diffcoeff}  
\usepackage{amsfonts,amssymb,amsmath}
\usepackage{mathtools}
\usepackage{comment}

\usepackage{bm}
\usepackage{times}
\usepackage{bbm} 
\usepackage{units}
\usepackage{mathrsfs}  

\usepackage{array} 
\usepackage{tabularx}
\usepackage{multirow}
\usepackage[caption=false]{subfig}			

\usepackage{hyperref}
\hypersetup{colorlinks=true,linktoc=all,linkcolor=blue,breaklinks=true,citecolor=blue,urlcolor=blue}

\usepackage[normalem]{ulem}  
\usepackage{xspace}
\usepackage[dvipsnames,table]{xcolor}
\usepackage{dsfont}

\addto\captionsenglish{}

\AtBeginDocument{%
    \newwrite\bibnotes
    \def\bibnotesext{Notes.bib}
    \immediate\openout\bibnotes=\jobname\bibnotesext
    \immediate\write\bibnotes{@CONTROL{REVTEX42Control}}
    \immediate\write\bibnotes{@CONTROL{%
            apsrev42Control,author="08",editor="1",pages="1",title="0",year="1"}}
    \if@filesw
    \immediate\write\@auxout{\string\citation{apsrev42Control}}%
    \fi
}%

\newcommand{\Tcol}[0]{T_\mathrm{c}}
\newcommand{\Tabs}[0]{T_\mathrm{h}}
\newcommand{\Tabszero}[0]{T_\mathrm{h,0}}
\newcommand{\Tabsone}[0]{T_\mathrm{h,1}}
\newcommand{\mucol}[0]{\mu_\mathrm{c}}
\newcommand{\mucolzero}[0]{\mu_\mathrm{c,e}}
\newcommand{\mucolone}[0]{\mu_\mathrm{c,f}}
\newcommand{\mucolstop}[0]{\mu_\mathrm{c}^\mathrm{stop}}
\newcommand{\muabs}[0]{\mu_\mathrm{h}}
\newcommand{\fcol}[0]{f_\mathrm{c}}
\newcommand{\fcolempty}[0]{f_\mathrm{c,e}}
\newcommand{\fcoli}[0]{f_{\mathrm{c},i}}
\newcommand{\fcolfull}[0]{f_\mathrm{c,f}}
\newcommand{\fabs}[0]{f_\mathrm{h}}
\newcommand{\pempty}[0]{p_\mathrm{e}}
\newcommand{\pfull}[0]{p_\mathrm{f}}
\newcommand{\avgpempty}[0]{\bar p_\mathrm{e}}
\newcommand{\avgpfull}[0]{\bar p_\mathrm{f}}
\newcommand{\ratetofull}[0]{\ensuremath{\Gamma_{\mathrm{fe}}}}
\newcommand{\ratetoempty}[0]{\ensuremath{\Gamma_{\mathrm{ef}}}}
\newcommand{\tcharge}[0]{\tau_\mathrm{charge}}
\newcommand{\tconv}[0]{\tau_\mathrm{conv}}

\newcommand{\kB}[0]{k_\mathrm{B}}
\newcommand{\edot}[0]{\varepsilon_\mathrm{QD}}
\newcommand{\en}[0]{\varepsilon}
\newcommand{\enone}[0]{\en_\mathrm{e}}
\newcommand{\entwo}[0]{\en_\mathrm{f}}
\newcommand{\D}[0]{D}
\newcommand{\enfilter}[0]{\en_\mathrm{filter}}

\begin{document}
    
\title{An autonomous feedback protocol: responding to temperature and potential changes in energy converters}

\author{Elsa Danielsson}
\thanks{These authors contributed equally}
\affiliation{Department of Microtechnology and Nanoscience (MC2), Chalmers University of Technology, S-412 96 G\"oteborg, Sweden\looseness=-1}

\author{Krishna Lyn Delima}
\thanks{These authors contributed equally}
\affiliation{Department of Microtechnology and Nanoscience (MC2), Chalmers University of Technology, S-412 96 G\"oteborg, Sweden\looseness=-1}

\author{Bruno Bertin-Johannet}
\affiliation{Department of Microtechnology and Nanoscience (MC2), Chalmers University of Technology, S-412 96 G\"oteborg, Sweden\looseness=-1}

\author{Janine Splettstoesser}
\affiliation{Department of Microtechnology and Nanoscience (MC2), Chalmers University of Technology, S-412 96 G\"oteborg, Sweden\looseness=-1}

\date{\today}

\begin{abstract} 

Nanoscale thermoelectric devices convert tiny amounts of heat into power. But what happens if the external conditions and resulting temperature differences are not \textit{a priori} known? We propose a feedback mechanism that optimizes the performance of a quantum-point-contact-based heat engine (QPC). A quantum-dot detector measures the external conditions and gives autonomous feedback on the QPC’s properties. We expect this feedback mechanism to be experimentally relevant for steady-state engines with unknown external conditions and for adapting dynamic charging processes to the potential buildup.

\end{abstract}

\maketitle

\textit{Introduction---} 
Is it possible for a quantum heat engine to adapt to its available resources without external control? Reactivity, responsiveness, and dynamics are key for many physical applications, but are not typically associated with thermoelectric devices which exploit \textit{steady-state} currents~\cite{Benenti2017Jun,Campbell2026Jan,Balduque2026Jan}. Nevertheless, steady-state quantum heat engines can be particularly beneficial since they avoid moving parts that can be difficult to engineer or control. 
Thermoelectric effects at the nanoscale were experimentally achieved in quantum point contacts some decades ago~\cite{vanHouten1992Mar}. More recently, it has been demonstrated that conversion of heat to work with quantum-dot devices can be implemented with high conversion efficiencies~\cite{Josefsson2018Oct,Volosheniuk2026Jan,Volosheniuk2025Apr}. Energy-harvesting has also been realized in multi-terminal devices with multiple capacitively coupled quantum dots~\cite{Thierschmann2015Oct,Jaliel2019Sep,Roche2015Apr}. In these implementations, the heat engines exploited temperature biases that were purposefully created by electrical heating. However, the conditions around a nanoelectronic device can be highly variable or not precisely known in practice. Examples are hot-carrier photovoltaics~\cite{Ross1982,Fast2021,Chen2020} where the external illumination can change; and distributions of quasiparticle excitations arising in small-scale quantum devices due to nearby operations, where the detection of temperature gradients is challenging~\cite{Krinner2019Dec,Nguyen2013Aug}.
An energy-converting device designed to work optimally for a certain temperature or potential bias may then have a significantly lower effect due to deviations from the optimal condition. It is therefore desirable to engineer the device to dynamically adapt to the available resources to improve the energy-conversion process.

\begin{figure}[tb]
    \subfloat{
      \includegraphics[width=\linewidth]{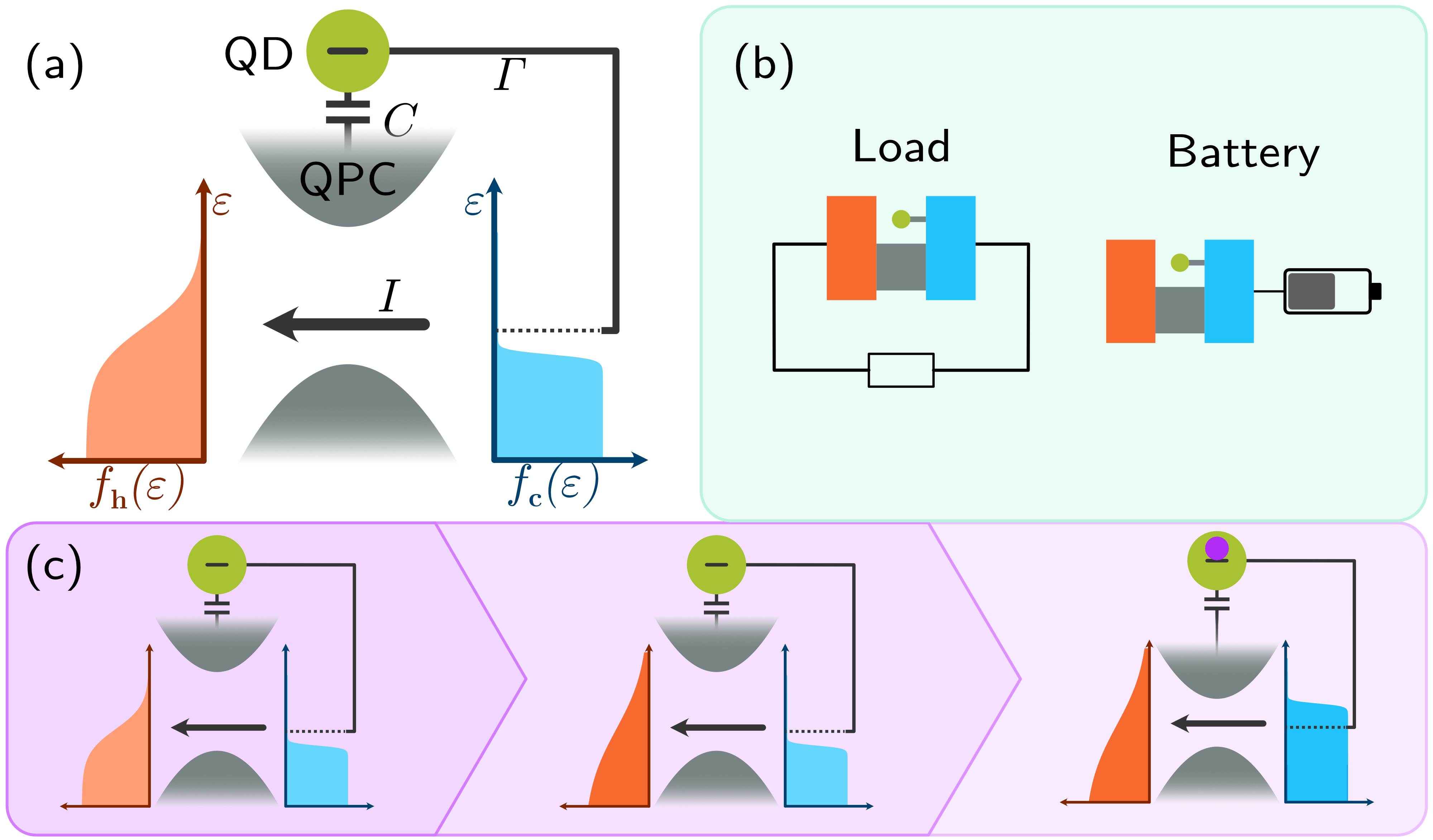}}
      \caption{Schematic of the autonomous feedback device. (a) A quantum point contact (QPC) conducts a (negative) charge current from a cold contact to a hot contact. The cold contact is coupled with strength $\Gamma$ to a quantum dot (QD), in turn capacitively coupled to the QPC with capacitance $C$. (b) The two circuit configurations investigated in the paper. The load setup connects the two contacts through an external resistance. The battery setup lets the cold contact act as a simple battery through charge build-up. (c) Conceptual process of the feedback mechanism. Initially the quantum dot is empty. Then the temperature is raised in the hot contact. The quantum dot is filled either by fluctuation or increased potential, narrowing the QPC's potential.} \label{fig:setup}
\end{figure}

In this Letter, we address this challenge and suggest an autonomous feedback protocol~\cite{Brandes2010,Emary2016} that adjusts the thermoelectric properties of a \textit{steady-state} energy converter in response to a variable temperature bias across the device. 
We consider a quantum point contact (QPC) acting as thermoelectric conductor between a hot input contact, providing heat as the resource, and a cold output contact in which the electrochemical potential rises as a result of the thermoelectric effect and can be exploited to provide electrical power~\cite{vanHouten1992Mar,Proetto1991Oct,Bogachek1998Nov,Dzurak1993Oct,Kheradsoud2019Aug}, see Fig.~\ref{fig:setup}. QPCs are the coherent conductors that yield maximum output power, given the ideal properties at a certain temperature and potential bias~\cite{Whitney2013Mar,Kheradsoud2019Aug,Benenti2017Jun}. However, when the temperature of the input contact or the potential of the output contact changes due to the external conditions, also the requirements on the ideal thermoelectric properties of the QPC for power production change. 
Here, we propose to couple to the cold output contact (assumed at base temperature) a quantum dot (QD) that reads out the electrochemical potential arising as the result of the (unknown) temperature bias. The QD is further capacitively coupled to the QPC, consequently providing autonomous feedback: when the dot occupation changes as the result of an altered temperature difference, the potential landscape in the QPC is modified, thereby improving the thermoelectric energy conversion. 
The choice to attach the quantum dot to the output and not to the input/resource contact is crucial to avoid the measurement precision being impacted by the high temperature smearing the electronic distribution in the hot resource. This choice enables a second use-case for our proposed feedback mechanism: improving the \textit{dynamic charging process} of a battery by the device adapting to the build-up of electrochemical potential. 

We therefore show the advantage of this feedback protocol with two relevant examples, see Fig~\ref{fig:setup}(b): (i) a battery (open circuit) setup, where feedback dynamically improves the charging process and (ii) a steady-state load (closed circuit) setup, where feedback improves the output power even if the temperature bias is only known at the level of a probability distribution. 

Feedback of a quantum dot onto the transmission of a QPC has been extensively studied in the context of single-charge detection~\cite{Lu2003May,Fujisawa2006,Gustavsson2006,Schleser2007,Ihn2009Sep,Guttinger2011Apr,Volk2013Apr,Garreis2023Jan} and thermometry~\cite{Gasparinetti2012Jun,Torresani2013Dec,Mavalankar2013Sep,Yang2019}. However, to our knowledge, this mechanism has not been used to boost the operation of thermoelectric devices, as we propose here. Feedback in thermoelectric devices has been widely investigated in recent years for the realization of information-driven engines~\cite{Campbell2026Jan,Goerlich2025Jan, Koski2015}, where feedback is typically employed based on the state of the conductor with the goal of rectifying fluctuations. The mechanism we suggest here is much simpler: the measurement dot directly reacts to the distribution of the contacts and hence to charge build-up or the provided resource. This makes the mechanism of practical relevance for energy-converting nanoscale devices.

\textit{Model and theoretical approach---} 
We model coherent steady-state thermoelectric transport through the QPC with scattering theory~\cite{Blanter2000Sep,Moskalets2011Sep, Buttiker1990, Landauer1957}, which is valid as long as electron interactions can be described at the mean-field level. 
By contrast, we use a rate equation for the stochastic dynamics of the QD occupation~\cite{Donarini}. 

The two electronic contacts, $\alpha=\text{h,c}$ (hot/cold) are characterized by Fermi functions, $f_\alpha(\en) = [1+ \exp[(\en-\mu_\alpha)/k_\mathrm{B}T_\alpha]]^{-1}$, where $\en$ is the particle energy and  $k_\mathrm{B}$  the Boltzmann constant.
The  hotter (input) contact, from which heat is extracted, has temperature $\Tabs$, while its electrochemical potential is set to ground, $\muabs \equiv 0$. 
Conversely, the temperature of the output contact is colder, $\Tcol<\Tabs$, and its electrochemical potential $\mucol$ can vary in response to thermoelectrically induced charge transfer. 

The central coherent conductor is represented by a transmission probability $D(\en)$, which  for a  QPC is given by~\cite{Buttiker1990Apr}
$
    \D(\en) = ({1+\exp[{(-\en+\enfilter)/\omega}])^{-1}}.
$
Here, $\enfilter$ is the energy around which the transmission changes and $\omega$ determines the width of the slope. We focus here on a sharp step function, $\omega/k_\mathrm{B}\Tcol \to 0$ and discuss the effect of $\omega\neq 0$ in the End Matter, as only amplitudes of numerical values are affected. 
The charge current into the output contact is then given by~\cite{Moskalets2011Sep}
\begin{equation}\label{eq:current}
        I = \frac{e}{h}\int_{-\infty}^\infty d\en \D(\en)[\fabs(\en)-\fcol(\en)],
\end{equation}
with the negative electron charge $e$ and Planck's constant $h$.

The core addition to the thermoelectric setup is the feedback mechanism, see Fig.~\ref{fig:setup}, enabled by a single-level quantum dot (QD) with energy level $\edot$. The QD is tunnel-coupled to the cold contact and capacitively coupled to the QPC. We assume the on-site interaction to be the largest energy scale, excluding double occupation and neglect spin, which would otherwise lead to a numerical factor in the QD tunneling rates~\cite{Splettstoesser2010Apr}; also, we assume that the QD's occupation impacts the QPC transmission, but that there is no feedback backaction, meaning that the occupation does not depend on the QPC state, justified by the negligible charge pile-up at the QPC \cite{Garreis2023Jan}. The QD's occupation is hence solely determined by the electronic distribution of the cold contact, while the QPC transmission, namely the energy filter $\enfilter$, is altered by the QD occupation. This feedback effect is illustrated in Fig.~\ref{fig:setup}(c) as the occupied QD constricting the QPC potential.  When the QD is empty $\enfilter = \enone$, determined by the physical design of the QPC, and when it is filled $\enfilter = \enone + U\equiv \entwo$, where $U = e^2/C$ with $C$ as the capacitance between QD and QPC. 

Furthermore,  we assume that electrons traverse the QPC on a much shorter timescale than the one at which the QD occupation changes, see for example experiments where a QPC is used to trace the occupation of a  capacitively coupled quantum dot~\cite{Lu2003May,Fujisawa2006,Gustavsson2006,Schleser2007,Ihn2009Sep,Guttinger2011Apr,Volk2013Apr,Garreis2023Jan}.  Therefore, they either ``see'' an empty \textit{or} a filled QD during the traversal; these two cases happen with probability $\pempty$ or $\pfull = 1-\pempty$.
The average current through the QPC hence depends on the average occupation of the QD.
\begin{eqnarray}\label{eq:tot_current}
    I &=& \pfull I_\mathrm{f} + \pempty I_\mathrm{e},   
\end{eqnarray}    
where $I_\mathrm{e},I_\mathrm{f}$ are the currents under the condition that the QD is empty or filled. The probabilities are determined by a rate equation
\begin{equation}\label{eq:rate-equation}
    \frac{d\pfull}{dt}  = \ratetofull(\edot) \pempty (t) - \ratetoempty(\edot) \pfull (t).  
\end{equation}
The rates $\ratetofull(\edot)$, empty to filled, and $\ratetoempty(\edot)$, filled to empty, are proportional to the tunnel-coupling strength $\Gamma$ between the QD and the output contact and are treated differently for the battery and load setups due to the differing relevant timescales.
 
\textit{Timescales and validity of approximations---} 
The theoretical description introduced above relies on a number of assumptions concerning the timescales of the device's dynamics. The fastest process is the feedback from the QD on the QPC transmission which is instantaneous, since it is due to an EM-field. We further assume thermalization in the contacts to be a much faster process than any electron-transfer process in the system, meaning that temperatures and electrochemical potentials in the contacts are always well-defined. 
First, this means that the tunneling dynamics of the QD are much slower such that its occupation probabilities can be determined by the rate equation, Eq.~\eqref{eq:rate-equation}. Second, it means that the charge build-up due to the thermoelectric current directly translates into a change in electrochemical potential in the cold output contact. The current through the QPC does not impact the temperature of the contacts due to their large heat capacitance~\cite{Jezouin2013Oct}.

The important distinction between the battery and the load setups discussed below are the timescales at which the QD detector records changes of the electrochemical potential in the cold contact. Concretely, in the battery setup the goal is to optimize the charging \textit{process}. The feedback hence needs to be active during the charging, meaning that we are interested in implementing a detector dot with tunneling that is faster than the build-up time of $\mucol$. By contrast, in the load setup, we aim to optimize the \textit{average} output power for occasionally changing temperature conditions. Therefore, $\mucol(t)$ can change much faster than $1/\Gamma$, effectively letting $\mucol(t)$ change instantaneously after an electron has jumped into or out of the dot. These two limiting cases are captured by approximations on the time-dependence of the rates $\ratetoempty$ and $\ratetofull$, either evaluating them for a frozen $\mucol$ in the first case or at the two different $\mucolzero$ and $\mucolone$ in the second. The timescales are discussed further in the End Matter and Supplemental Material.

\textit{Open circuit (``battery") setup---}
In this setup, the cold output contact is assumed to be a \textit{finite} metallic island~\cite{Schaller2017,vandenBerg2015Jul,Pekola2013Oct}. The goal is to charge the metallic island---namely to increase its electrochemical potential---exploiting the thermoelectric current induced by a given temperature difference $\Tabs>\Tcol$. 

The idea is similar to Ref.~\cite{Schaller2017}, where an active feedback protocol was proposed to improve the charging process, in contrast to our \textit{autonomous} feedback scheme. The battery setup is the first scheme we discuss in detail, since the fixed resource temperature makes the problem conceptually simpler.
 
We start by introducing the charging dynamics in the absence of feedback. The induced charge current into the metallic island results in an increase of its electrochemical potential, which in turn counteracts the current flow
\begin{equation}\label{eq:capacitor_current}
    I(\mucol) = \frac{dq}{dt} = en_\mathrm{c}\frac{d\mucol}{dt},
\end{equation}
where $q$ is the charge accumulating on the metallic island. Here we have assumed the wide-band limit and a band-bottom far below the electrochemical potential to evaluate $dq/d\mucol = en_\mathrm{c}$, with $n_\mathrm{c}$ the number of available electronic states per energy. The current $I(\mucol)$ corresponds to Eq.~\eqref{eq:current}, which instantaneously depends on time through the time-dependent electrochemical potential $\mucol$.  

\begin{figure}[tb]
        \includegraphics[width = \columnwidth]{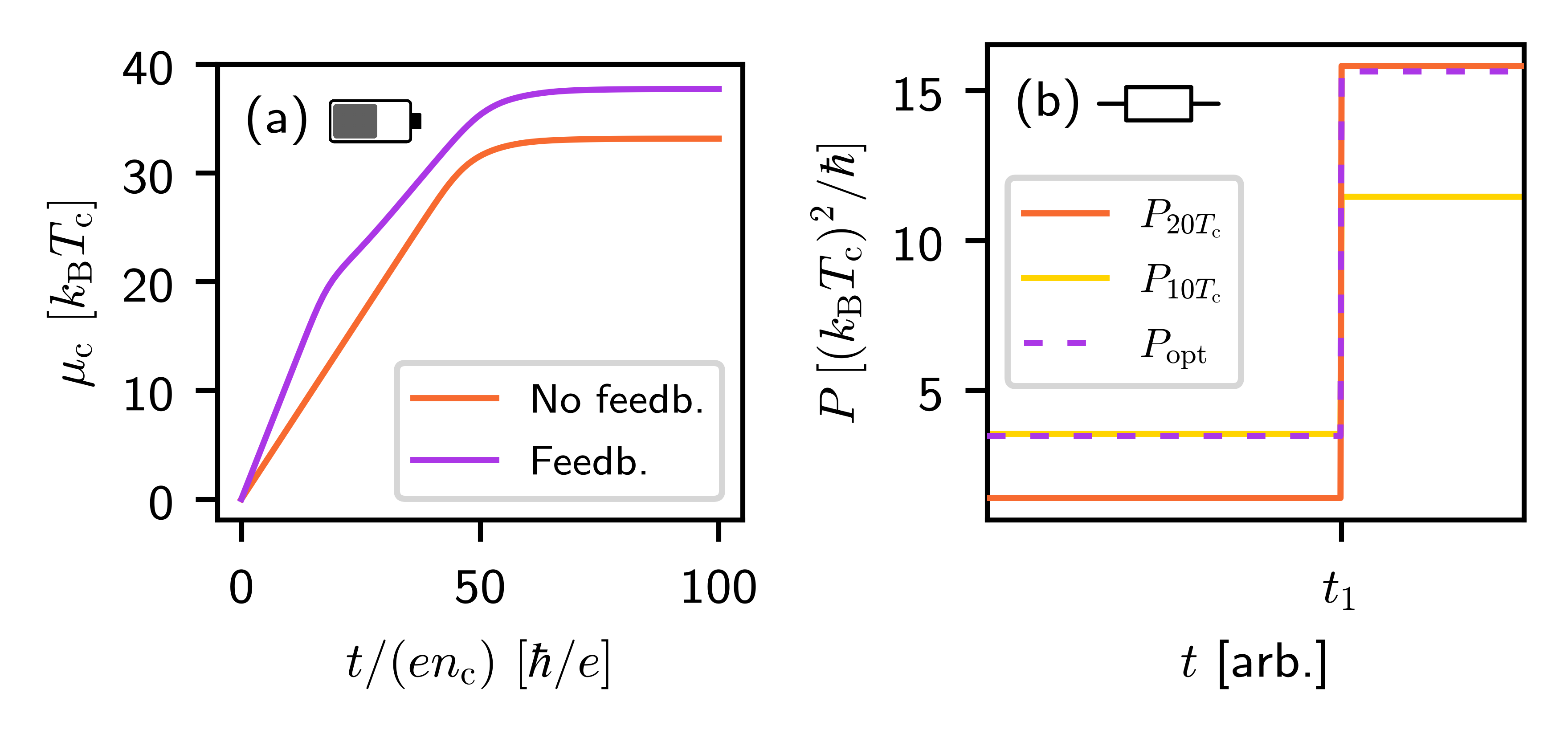}
    \caption{(a) Expected build-up of $\mucol$ in battery setup, obtained from Eq.~\eqref{eq:mu_ode}, for maximized $\mucol$ at $\tcharge/(en_\mathrm{c})= 50\ \hbar/e$ with $\Tabs = 20\ \Tcol$. The optimized parameters are $\enfilter = 28.95 \ \kB \Tcol$ for no feedback (red line) and $\enone = 17.62 \ \kB\Tcol$, $\entwo = 34.49 \ \kB\Tcol$, and $\edot = 20.45 \ \kB\Tcol$ with feedback (purple line). (b) Power outputs of the load setup at temperatures $\Tabs = 10\Tcol$ for $t<t_1$, and $\Tabs = 20 \Tcol$ for $t>t_1$, with $R = 1.6 \ h/(e^2)$. The filter heights are $\enone =  6.62 \ \kB \Tcol$ and $\entwo = 13.24 \ \kB \Tcol$, which are the optimal heights for $\Tabs = 10\Tcol$ and $\Tabs = 20\Tcol$ respectively, with corresponding powers in yellow and red. The optimized dot level for the purple feedback line is $\edot = 9.27 \ \kB\Tcol$.} 
    \label{fig:ideal_result} 
\end{figure}

Equation~\eqref{eq:capacitor_current} hence provides a self-consistent equation for the time-dependent buildup of the island potential $\mucol$ until it reaches $\mucolstop$; the stopping voltage is the value of $\mucol$ at which the current flowing into the island is canceled by the backflow due to the potential difference and its value depends on the choice of the energy filter, $\enfilter$. With a sufficiently large $\enfilter$, the potential $\mucol$ could reach any arbitrarily high value---but the charging time increases exponentially, due to the exponential decay of the Fermi function at high energies. We are therefore interested in the more practically relevant question of which stopping voltage can be reached in a given charging time. The red line in Fig.~\ref{fig:ideal_result}(a) shows how $\mucol$ changes over time in the absence of feedback, where the filter height $\enfilter$ is chosen to optimize $\mucol$ for $\Tabs=20\Tcol$ at the charging time $\tcharge/(en_\mathrm{c})= 50\ \hbar/e$. Notably, the potential $\mucol$ changes linearly until near $\mucolstop$. 

We now introduce feedback with the goal to increase the achievable $\mucolstop$ in the same fixed time interval $\tcharge$. The QD gets filled when $\mucol\gtrsim\edot$ is reached, consequently shifting the energy filter from height $\enone$ to $\entwo$. 
The rates determining the QD occupation are readily identified as
\begin{equation} \label{eq:battery_rates}
    \ratetoempty = \Gamma(1-\fcol(\edot)), \ \ratetofull = \Gamma\fcol(\edot),
\end{equation}
where the temporal trajectory of $\mucol(t)$ is accounted for in the Fermi functions. The rate equation Eq.~\eqref{eq:rate-equation} continues to be valid for a finite-sized reservoir given that the electron population thermalizes faster than the dot dynamics \cite{Moreira2023}.
The time-dependent current in the presence of feedback is now given by
\begin{equation}\label{eq:mu_ode}
    I(\mucol) = \pempty(\mucol) I_\mathrm{e}(\mucol) + \pfull(\mucol)I_\mathrm{f}(\mucol) = en_\mathrm{c} \frac{d\mucol}{dt},
\end{equation}
and is found by solving the differential equation~\eqref{eq:mu_ode} for $\mucol$ and the rate equation~\eqref{eq:rate-equation} simultaneously.\footnote{Practically, if $1/\Gamma$ is less than the timescale $\mucol(t)$ changes on, one could use the steady-state solutions $\avgpfull$ and $\avgpempty$ to only solve Eq.~\eqref{eq:mu_ode} due to the small effect of the decay, see the Supplemental Material.} This yields an expected temporal trajectory for $\mucol(t)$, averaging out the discrete jumps of the dot occupation. The trajectory is physically meaningful for a chosen $1/\Gamma$ for the detector dot that is smaller than charging time $\tcharge$ determining the temporal change of $\mucol(t)$.

The purple line in Fig.~\ref{fig:ideal_result}(a) shows that a higher stopping voltage is achieved when the feedback is in use, where $\enone,\entwo=\enone+U$ and $\edot$ were optimized for the given temperature bias. This clearly demonstrates the benefit of feedback when the parameters for energy levels and capacitive coupling can be appropriately implemented in a device. 
 In contrast to the no-feedback line there are two distinct slopes, corresponding to the charge build-up during the beginning of the charging process when $\enfilter = \enone$ and during the time when $\enfilter = \entwo$ until the stopping voltage is reached. During these times the QD is nearly always empty or nearly always filled, thus while $\mucol$ is smaller than the corresponding filter height the current is almost independent of $\mucol$ .  
This linear approximation is detailed in the Supplemental Material along with further discussion on charging dynamics.  
  
\textit{Closed circuit (``load") setup---}
The closed circuit connects the two contacts through an external load with resistance $R$. We are here interested in a steady-state situation, where power $P= \mucol^2/(e^2R) = R I^2$ is produced due to  the thermoelectrically induced current against a potential bias $\mucol\equiv\mucol-\muabs$.\footnote{Note that one could alternatively design a device where feedback modifies the load $R$ instead of the filter height influencing the current $I$.}  Again starting from the no-feedback case, this potential bias has to be found self-consistently from $\mucol/e=RI(\mucol)$. Results for two temperature biases $\Tcol,\Tabs=\{10\Tcol,20\Tcol\}$ are shown as red and orange lines in Fig.~\ref{fig:ideal_result}(b). 

In the load setup, we want the feedback mechanism to adapt to different possible stationary values of $\Tabs$. This represents a situation where the device operates at one stationary temperature, but its value is unknown; or the temperature varies in time, but on a timescale much slower than the dynamics of the feedback mechanism. We start by the simple situation where the hot input contact can have either one of the two temperatures $\Tabszero,\Tabsone$, but it is not known which one. 
Assuming that the potential $\mucol$ changes \textit{immediately} whenever the occupation of the quantum dot changes, the potential fluctuates between two distinct values. The steady-state (time-averaged) current is a mixture of the currents for the fraction of time the dot is occupied and when it is not, evaluated by the average probability of the QD's occupation. At a given pair of temperatures $\Tabs,\Tcol$ the current $I = \avgpfull I_\mathrm{f} + \avgpempty I_\mathrm{e}$ is thus
\begin{eqnarray} 
    I =   \sum_{i=\mathrm{e,f}}\bar{p}_i\frac{ e}{h}\int_{\epsilon_i}^\infty d\en [\fabs(\en)-\fcoli(\en)] .     \label{eq:tot_current}
\end{eqnarray}   
Here, we have indicated that the Fermi function of the cold output contact is different depending on whether its potential is given by $\mucolzero$ resulting from the filter transmission with $\enfilter=\enone$ due to an empty QD or whether its potential is $\mucolone$ resulting from a filter transmission with $\enfilter=\entwo$ due to a filled QD. 
The steady-state probabilities $\avgpempty$  and $\avgpfull=1-\avgpempty$  are the steady-state solutions of Eq.~\eqref{eq:rate-equation} with
\begin{equation} \label{eq:load_rates}
    \ratetoempty(\edot) = \Gamma(1-\fcolfull(\edot)), \ \ratetofull(\edot) = \Gamma\fcolempty(\edot).
\end{equation}
The rate going from a full to an empty dot is determined by the distribution in the cold contact with $\mucolone$, and similarly for the rate empty to full. In contrast to the rates for the battery setup, Eq.~\eqref{eq:battery_rates}, these rates depend on different electron distributions that are \textit{fixed} for fixed filter heights. Consequently, the feedback mechanism relies on fluctuations in the quantum-dot occupation, which require finite rates in both directions. The magnitude of the rates affect the timescale for the exponential decay to the steady-state solution to Eq.~\eqref{eq:rate-equation}, imposing a timescale limitation for reaching the steady-state current, Eq~\eqref{eq:tot_current}, and on temperature changes, which we discuss in the End Matter. 

We are interested in optimizing the average power output, which at a given temperature $\Tabs$ is
\begin{equation} \label{eq:power_mixture}
    P = \avgpfull \frac{\mucolone^{2}}{eR} + \avgpempty \frac{\mucolzero^{2}}{eR}
\end{equation}
By choosing $U$, $\enone$ and $\edot$, we can create a device that operates \textit{ideally} at two distinct temperatures, $\Tabszero,\Tabsone$ as illustrated in Fig.~\ref{fig:ideal_result}(b). If $\Tabs \gg \Tcol$, the potential difference $\Delta \mucol = \mucolone(\Tabsone)-\mucolzero(\Tabszero)$ is much greater than $\kB \Tcol$. This means that if $\edot$ is put in between $\mucolone(\Tabsone)$ and $\mucolzero(\Tabszero)$, the dot is almost completely empty when $\Tabs = \Tabszero$ or completely full when $\Tabs = \Tabsone$. Thus $\enone$ can be optimized for the lower $\Tabszero$ and $\entwo$ for the higher $\Tabsone$ such that the device yields the highest possible output power at both temperatures. This is shown in Fig.~\ref{fig:ideal_result}(b), where we assume that the external conditions change from $\Tabszero$ to $\Tabsone$ at an arbitrarily chosen time $t_1$.
Without the feedback, the filter height is optimized for one pair of temperatures $\Tabs,\Tcol$, but when the hot temperature changes the power output is less than ideal. By contrast, with the feedback, the filter changes between the two heights, giving in total a higher power output. 

\begin{figure}[tbh] 
        \includegraphics[width = \columnwidth]{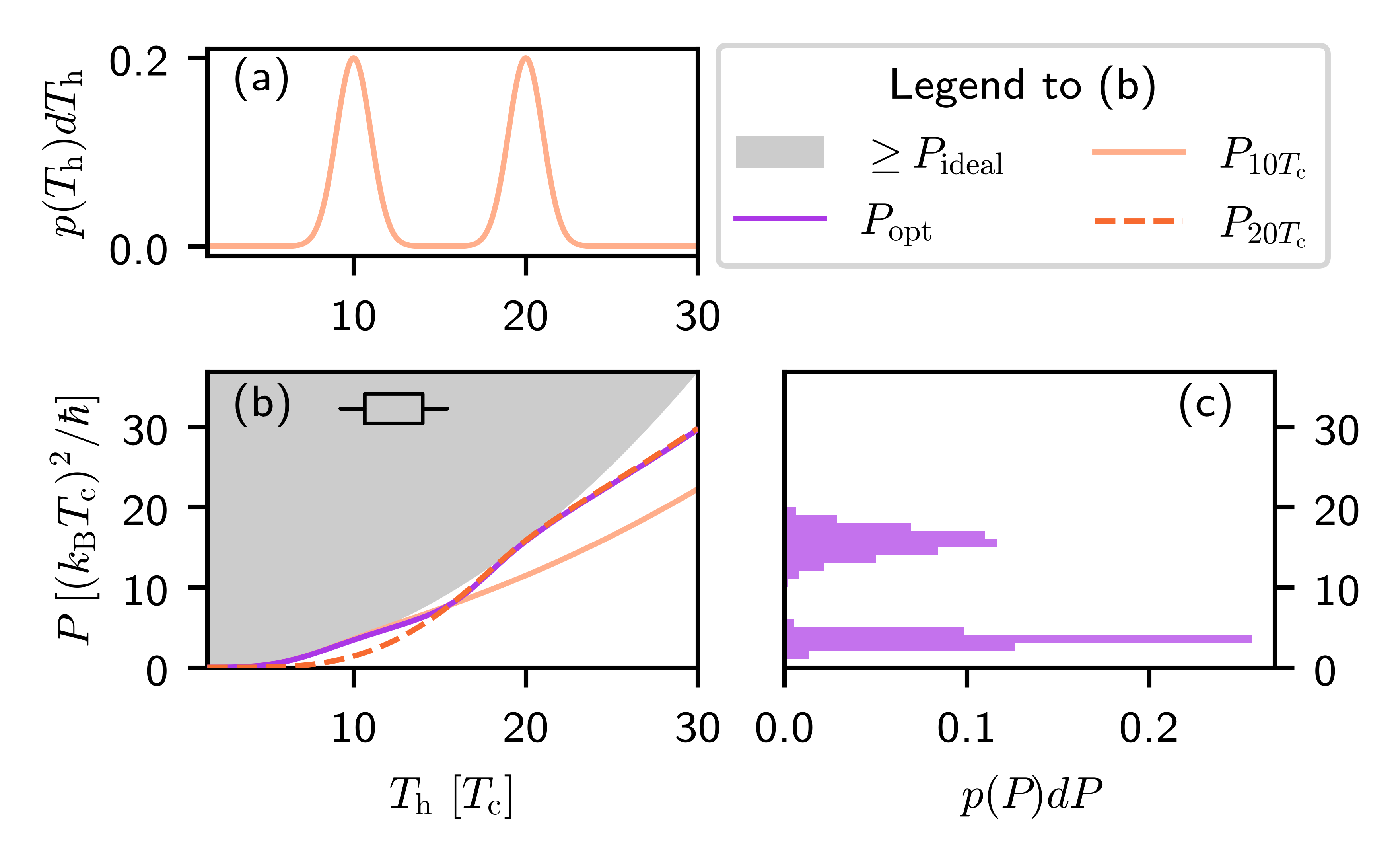}
    \caption{ (a)~Bimodal probability distribution for $\Tabs$, constructed by normal distributions centered at $ 10 \Tcol$ and $ 20 \Tcol$ with variance $\sigma^2 =  \Tcol$. (b)~Produced power in the load setup: $P_\mathrm{ideal}$ is  achieved when the barrier height is optimal at each temperature (all values in the gray region are hence excluded); $P_{10\Tcol}$ and $P_{20\Tcol}$ are the powers when the barrier height is optimal for $\Tabs = 10 \ \Tcol$ and $\Tabs = 20 \ \Tcol$ respectively, see Fig.~\ref{fig:ideal_result}; $P_\mathrm{opt}$ is the power for feedback with filter heights optimized for maximum average power given the bimodal distribution of (a): $\enone = 6.68 \ \kB\Tcol$, $\entwo = 13.10 \ \kB\Tcol$, $\edot = 8.73\ \kB\Tcol$. (c) Probability distribution of the power values of $P_\mathrm{opt}$, see Eqs.~\eqref{eq:average_power}.} 
    \label{fig:temp_dist_results}
\end{figure}
\textit{Temperature distributions---} 
Hitherto, the load setup operated at one of two possible pairs of temperatures $\Tabszero,\Tcol$ and $\Tabsone,\Tcol$. Under these simplified conditions, ideal feedback parameters $\enone,U,\edot$ were identified to optimize the device operation.  But what if the range of possible hot temperatures of the input contact is not \textit{a priori} known? It is then not possible to adapt $\enone,U,\edot$ to each individual external condition, but it is rather desirable to identify parameters that yield an optimized operation for a \textit{distribution} of possible hot temperatures $\Tabs$.  
The goal is to maximize the mean power output $\bar{P}$ (load setup). We hence want to optimize a stochastic variable $X$ ($P$ here and furthermore $\mucol$ in the Supplemental Material) with mean value 
\begin{equation} \label{eq:average_power}
    \bar{X} = \int Xp(X)dX  = \int Xp(\Tabs(X))\frac{d\Tabs(X)}{dX} dX.
\end{equation}
where $p(X)dX$ is the probability distribution for $X$ found through a change of variables \cite{Arfken}, see Ref.~\cite{zenodo} for numerical details and data. 
We achieve the maximum mean value $\bar{P}$ by optimizing $\enone$, $\entwo=\enone+U$, and $\edot$ for a given probability distribution $p(\Tabs)$.

The results for a bimodal temperature distribution are shown in Fig.~\ref{fig:temp_dist_results}. With a bimodal distribution, we represent a situation where temperature regions around two external conditions are most probable (here with normal distributions centered around $\Tabs = 10\Tcol$ and $\Tabs = 20\Tcol$ with $\sigma = 1 \ \Tcol$, see panel~(a)). Other viable temperature distributions are shown for reference in the Supplemental Material. 
The border of the gray area in (b) represents the  bound $P_\mathrm{ideal}$, calculated by individually setting the ideal filter height at each temperature $\Tabs$ (which is practically not realizable). The red lines show the power obtained at each temperature $\Tabs$ for no-feedback devices where the filter is chosen such that it is ideal for $\Tabs=10\Tcol$, or for  $\Tabs=20\Tcol$ for the dashed line, and they therefore saturate the bound at one point. The power $P_\mathrm{opt}(\Tabs)$, shown in purple, is obtained for a feedback device using parameters optimized to get highest \textit{average} power output for the bimodal distribution. The curve saturates at two points and is always closest to the higher of the two red curves. This demonstrates the benefit of the additional degree of freedom introduced by the feedback mechanism

The corresponding $p(P)dP$ (found through histogramming $P_\mathrm{opt}$ \cite{zenodo}) is shown in Fig.~\ref{fig:temp_dist_results}(c), clearly reflecting the two bumps in the power curve as peaks in the probability distribution. Using the distribution in Eq.~\ref{eq:average_power}, the average power for the feedback curve is $\bar P_\mathrm{opt} = 9.53 \ (\kB \Tcol)^2/\hbar$, compared to $\bar P_{20 \Tcol} = 8.62 \ (\kB \Tcol)^2/\hbar$ and $\bar P_{10 \Tcol} = 7.51 \ (\kB \Tcol)^2/\hbar$. Thus the feedback mechanism allows for a larger average power output (here by more than 10\%) given \textit{one set} of well-tuned parameters.  

\textit{Conclusion---}
We have introduced an autonomous feedback protocol that significantly increases the performance of thermoelectric devices by autonomously adapting their working principle to external conditions. 
We have demonstrated the applicability of our feedback scheme both for an increased charging potential of a battery setup in a given charging time and for a load setup adapting to different temperature biases. Importantly our approach is beneficial even if the external conditions (namely the temperatures of the heat bath) are known only up to a probability distribution of the temperature.
We predict that the feedback mechanism could be useful to exploit thermal resources on-chip in practical situations where the working conditions depend on external factors. Indeed, experimental platforms using the same feedback principle already exist and are hence in experimental reach. 
In the future, it would be intriguing to analyze additional opportunities that occur when using continuous charge or multilevel dots as detectors.

\acknowledgements
We thank Gernot Schaller and Kay Brandner for insightful discussions, and Sofia Sevitz and Ludovico Tesser for helpful comments on the manuscript. Funding from the Knut and Alice Wallenberg Foundation through the Fellowship program and from the European Research Council (ERC) under the Horizon Europe research and innovation program of the European Union (101088169/NanoRecycle) is gratefully acknowledged.

\bibliography{refs}

\section{END MATTER}
\subsection{Estimating timescales}\label{EM:convergence}
In this Section we give an account for the timescales highlighted in Table~\ref{tab:timescales}. 
\begin{table}[tbh]
    \centering    
    \caption{Relevant timescales for the device as order-of-magnitude estimates. Ordered top-down from slowest to fastest. Timescales highlighted in pink are estimates based on parameters used in this paper.} 
    \begin{tabular}{|l|c|c|}
    \hline
        \textbf{Process} & \textbf{Parameter} & \textbf{Timescale} \\ \hline
        Temperature change & - & seconds or more  \\
        \rowcolor{pink}
        Load steady-state & $\Gamma, \eta, U$ & $ \tconv \sim 10/\Gamma-100/\Gamma$ \\
        \rowcolor{pink}
        Charge build-up & $\enone, \entwo, \Tabs, n_\mathrm{c}$ & $\tau_\mathrm{charge} \sim 1-10$ ns  \\
        QD jump frequency & $\Gamma$ &  $1/\Gamma \sim$ ns-ms \cite{Lu2003May, Gustavsson2006, Li2013Oct, Goldhaber-Gordon1998Jan} \\
        QPC traversal time & $\omega$ (width) & ps \cite{Wolf2009}\footnote{Traversal time estimated from $\omega/v_\mathrm{F} \sim \frac{100 \ \mathrm{nm}}{10^{5} \ \mathrm{m/s}} = 1 \ \mathrm{ps}$} \\
        Thermalization & - & fs \cite{Pothier1997Nov, Roulet2024}  \\ \hline
    \end{tabular}
    \label{tab:timescales}
\end{table}
\subsubsection{Battery setup}
The charge build-up time, the second highlighted row in Table~\ref{tab:timescales}, is defined as the time it takes for the cold contact (the battery) to charge from $\mucol = 0$ to $\mucolstop$. This charging time $\tcharge$ is comparable to the $RC$-time of an $RC$-circuit, keeping in mind that charge and relaxation times in nonlinear and quantum circuits can differ from classical $RC$-circuit behavior \cite{Splettstoesser2010Apr,Kashuba2012Jun,Freitas2020Jul}. Here $\tcharge$ is identifiable from the feedback trajectory in Fig.~\ref{fig:ideal_result}(a). Since this particular trajectory is optimized with respect to a \textit{given} charging time $\tcharge/(en_\mathrm{c})= 50\ \hbar/e$, we can estimate its numerical value directly. With $\Tcol = 100 \ \mathrm{mK}$ and $en_\mathrm{c} = 1 \ e/(\kB\Tcol) $, we have $\tcharge \approx 4 \ \mathrm{ns}$, motivating the timescale in the table. For a general setting, the charging timescale grows directly with both $\Tcol$ and with the density of states via $en_\mathrm{c}$; smaller $\Tcol$ give larger $\tcharge$ and larger $en_\mathrm{c}$ give larger $\tcharge$. The charging time furthermore depends on the filter heights, which are the parameters we modulate in the main text to optimize the charging process. For a given set of parameters, $\tcharge$ is found by numerically solving the ODE~\ref{eq:mu_ode}, but can also be estimated through a linear approximation as detailed in the Supplemental Material. 
\subsubsection{Load setup}
Here we describe how the timescale in the first highlighted row in Tab.~\ref{tab:timescales} is estimated. This timescale corresponds to the time needed to for the feedback-enabled device to reach steady-state operation. This represents a situation where the device is initialized at a given, fixed $\Tabs$ with an empty quantum dot. Then through electrons jumping in and out of the QD, the device eventually reaches steady-state, which is when the rate equation~\eqref{eq:rate-equation} approaches the steady-state solution. At this point, the current in Eq.~\eqref{eq:tot_current} becomes meaningful as a statistical mixture between the current while the QD is empty and the current while it is filled.
 
From the rate equation~\eqref{eq:rate-equation} with time-independent rates~\eqref{eq:load_rates}, the probability $\pfull(t)$ rapidly decays exponentially towards $\avgpfull$. To fully reach the steady-state, we need to take $t\to\infty$, but in practice $\pfull(t)$ will be ``close enough'' to $\avgpfull$ at a finite time. To quantify ``close enough'', we define a tolerance $\eta$ as
\begin{equation}
    |\avgpfull - \pfull(\tau_\mathrm{conv})| = \eta,
\end{equation}
where $\tau_\mathrm{conv}$ is the convergence time determining the timescale for $\pfull(t)=1-\pempty(t)$. 

This time is bounded from above, which what is used to inform the timescale for reaching steady-state operation.

Since the rates, Eq.~\eqref{eq:load_rates}, depend on two different distributions, $\fcolempty(\en)$,$\fcolfull(\en)$, it is possible for the quantum-dot state to get ``stuck'' as empty or full although the average probability is between zero and one. This effect is most prominent if $\edot = \edot^\mathrm{max}\equiv(\mucolone + \mucolzero)/2$, for some $\Tabs$. With the solution of the rate equation this leads to the bound
\begin{equation}
    \tconv \leq -\frac{1}{\Gamma}\frac{1}{2\fcolempty((\mucolone + \mucolzero)/2)} \ln(\eta)
\end{equation}
This bound can furthermore be approximated using $U = \entwo-\enone$ to
\begin{equation} \label{eq:conv_with_U}
    \tconv \lesssim -\frac{1}{\Gamma} \frac{1}{2f_\mathrm{c,0}(\varphi U/4)}\ln(\eta),
\end{equation}
where $f_\mathrm{c,0}(\en)= [1+\exp(\en/\Tcol)]^{-1}$  and $\varphi = (Re^2/h)/(1+Re^2/h)$. From this formula, we see that the timescale is at least a few $1/\Gamma$ if $\eta$ is some reasonably small tolerance, $\sim 10^{-5}$ for instance, and that the timescale grows exponentially with $U$. 
Detailed derivations and discussions are found in the Supplemental Material, along with an upper bound on $U$. 

While the timescale evaluation and the current mixture in Eq.~\eqref{eq:tot_current} rely on a fixed $\Tabs$, the temperature could be allowed to change in time if there is a large timescale separation between the decay to the steady-state and the temperature changes in time. Then the device can be considered reinitialized at each new temperature and the probabilities then reach steady-state while $\Tabs$ remains effectively constant. Temperature change could also be rapid if it then remains constant for a longer time, as in Fig.~\ref{fig:ideal_result}(b). For the brief time the temperature switches there is no steady-state operation, but steady-state is reached for the temperatures separately while they are constant. 

\subsection{Smooth transmission function}\label{EM:smooth} 

Throughout the main text, the transmission function for the QPC is approximated as a sharp energy filter. Here, we demonstrate how a finite $\omega$ affects some of the results. Conceptually, the smearing of the transmission function will only affect numerical values of parameters while keeping the core principle of the feedback mechanism. 

\begin{figure}[tbh]
    \centering
    \includegraphics[width=\linewidth]{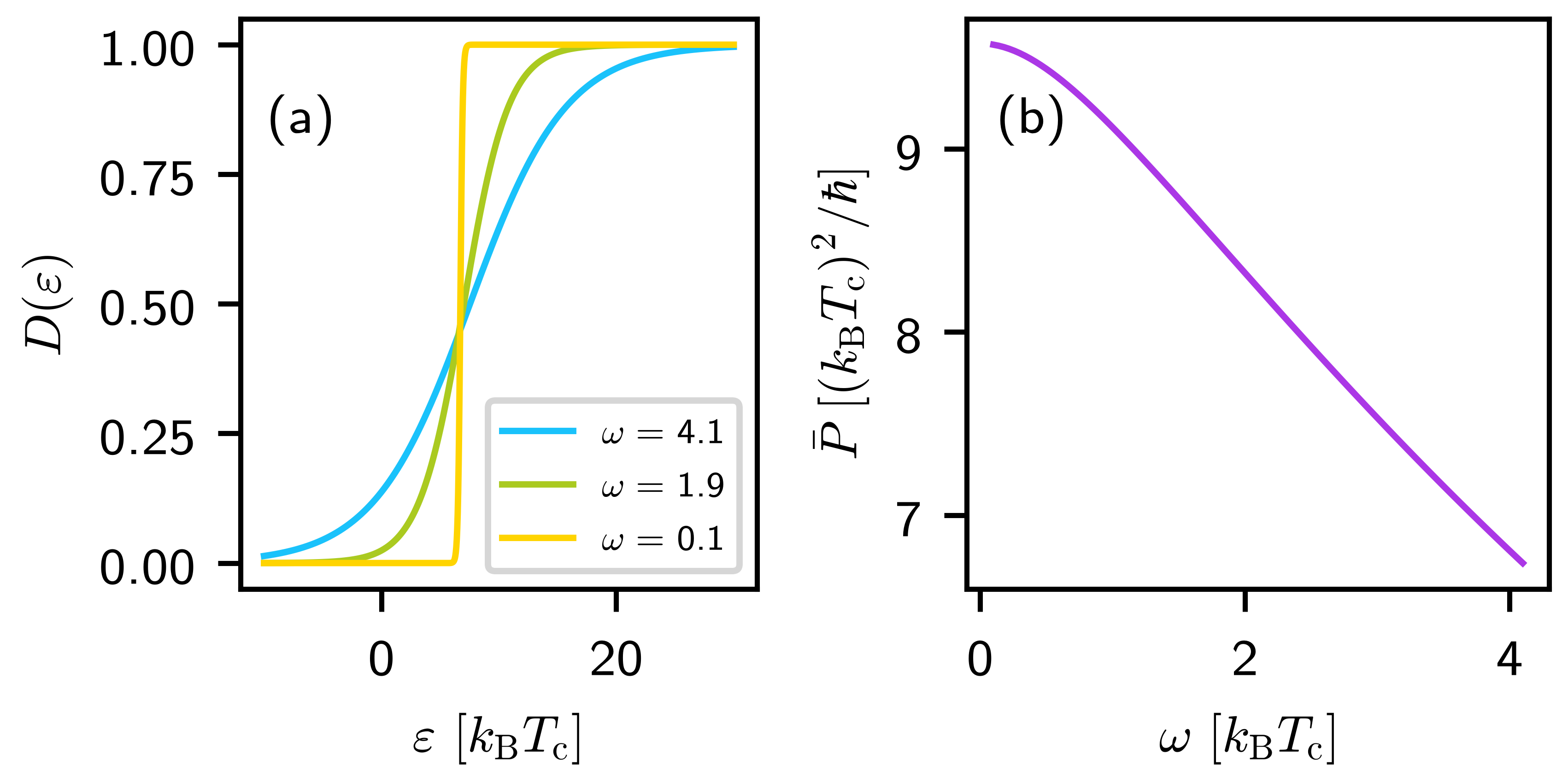}
    \caption{(a) Transmission functions for QPCs of various widths. (b) Maximized average power depending on width of the QPC transmission function. Parameters provided in Ref.~\cite{zenodo}.}
    \label{fig:smooth_p}
\end{figure}

We consider here the load setup with a smooth transmission function and two input temperatures $\Tabszero = 10\Tcol$ and $\Tabsone = 20\Tcol$. As a reference, transmission functions of different smoothness are plotted in Fig.~\ref{fig:smooth_p}(a). As a measure on the effect of the smoothness $\omega$, the power output averaged on the two temperatures is evaluated, with feedback parameters optimized for each $\omega$ to maximize this average power. Figure~\ref{fig:smooth_p}(b) shows this evaluation, where the averaged power clearly trends downward with increased smoothness.  

The maximum possible current at the two temperatures are found when the transmission function is a sharp energy filter, with parameters as in Fig.~\ref{fig:ideal_result}(b). For increasing $\omega$ the filter heights $\enone$ and $\entwo$ both increase in order to move the transmission window away from energies where electrons would transfer from the cold contact to the hot one. As the transmission function is no longer ideal, the current decreases and consequently the potentials $\mucolzero$ and $\mucolone$ decrease with increasing smoothness.

\end{document}


\title{Supplemental material to "An autonomous feedback protocol: responding to temperature and potential changes in energy converters"}

\author{Elsa Danielsson}
\thanks{These authors contributed equally}
\affiliation{Department of Microtechnology and Nanoscience (MC2), Chalmers University of Technology, S-412 96 G\"oteborg, Sweden\looseness=-1}

\author{Krishna Lyn Delima}
\thanks{These authors contributed equally}
\affiliation{Department of Microtechnology and Nanoscience (MC2), Chalmers University of Technology, S-412 96 G\"oteborg, Sweden\looseness=-1}

\author{Bruno Bertin-Johannet}
\affiliation{Department of Microtechnology and Nanoscience (MC2), Chalmers University of Technology, S-412 96 G\"oteborg, Sweden\looseness=-1}

\author{Janine Splettstoesser}
\affiliation{Department of Microtechnology and Nanoscience (MC2), Chalmers University of Technology, S-412 96 G\"oteborg, Sweden\looseness=-1}

\date{\today}

\maketitle

\section{Traversal and thermalization times}

In the main paper, we always assume that the thermalization time is the fastest and that the traversal time of electrons through the QPC is always much shorter than the timescale on which the detector dot state evolves,
\begin{equation}
    \tcharge,\frac{1}{\Gamma}\gg \text{QPC-traversal\ time}\gg \text{thermalization\ time},
\end{equation}
where $\tcharge$ is the charging time and $\Gamma$ is the coupling between the quantum dot and the cold contact. Consequently, the current flowing through the QPC takes different values depending on whether the dot is empty or filled:
\begin{subequations} \label{eq:empty_full_currents}
\begin{eqnarray}
    I_\mathrm{e}  &= \int_{\enone }^\infty d\en(\fabs(\en) - \fcol(\en, \mucol(t))),\\
    I_\mathrm{f}  &= \int_{\entwo }^\infty d\en(\fabs(\en) - \fcol(\en, \mucol(t))).    
\end{eqnarray}    
\end{subequations}
It is thanks to the fast thermalization time that there is a given value of $\mucol(t)$ at any moment in time, both in the battery and in the load setup. At timescales much larger than $1/\Gamma$ the current is hence a mixture of the two different values given in Eq.~\eqref{eq:empty_full_currents}
\begin{equation}\label{eq:current-mix}
    I(\mucol) = \pempty I_\mathrm{e}(\mucol) + \pfull I_\mathrm{f}(\mucol) .
\end{equation}
The properties of the probabilities $\pfull,\pempty$ depend on the setting (battery/load) we are considering. We are here considering weak coupling between detector dot and thermalized cold contact, $\Gamma\ll k_\mathrm{B}\Tcol$. Because the bath is thermalized between dot jumps, the quantum trajectory of the dot population is Markovian. Therefore the dynamics of the dot occupation can be described by a master equation~\cite{Benenti2017Jun}
\begin{equation} \label{eq:rate_equation}
    \frac{d\pfull}{dt}  = \ratetofull \pempty (t) - \ratetoempty \pfull (t),
\end{equation}
where $\pfull(t)$ represents the probability of finding the quantum dot occupied at time $t$ and $\pempty(t) =  1- \pfull(t)$ the probability of finding it empty. 

\section{Dynamics of $\mucol$ for the battery setup}
Here, we show details about how to evaluate the dynamic buildup of the potential $\mucol$ in the cold contact in the battery setup. We assume that the temperature of the resource contact is constant. Otherwise there would be a time-dependence in $\mucolzero$ and $\mucolone$, which would require both knowing how $\Tabs$ changes in time and solving a more complicated rate equation. 
The situation described here corresponds to a setting where a given hot contact is available and the ``battery" is attached to it for charging at the initial time $t=0$. 
 
 The continuous process for the feedback mechanism for the battery setup can be imagined as occurring in the following ``steps":
\begin{enumerate}
    \item An electron jumps into or out of the detector quantum dot on a timescale $1/\Gamma$
    \item The filter energy of the QPC transmission shifts up or down, instantaneously
    \item The potential $\mucol$ evolves according to  $I_\mathrm{e,f}= qn_\mathrm{c} \frac{d\mucol}{dt}$, where $\tcharge$ sets the timescale on which the evolution takes place 
    \item The process repeats when the quantum dot population changes again (influenced by the changes of the electrochemical potential $\mucol$). 
\end{enumerate}
The timescales determining the battery charging have the following relative magnitudes
\begin{equation}
    \tcharge\gg\frac{1}{\Gamma}
\end{equation}
meaning that we are interested in a feedback mechanism that acts along the charging process - the detection/feedback hence needs to be much faster than the charging.
This means that switches back and forth of the dot occupation (and consequently of the QPC transmission) happen on a much shorter timescale than the timescale of the potential evolution. By using the master equation~\eqref{eq:rate_equation} we average over these switches, which we are here not interested in, thereby focusing on the average dynamics.

The rate of change for $\mucol$ then depends on the mixture of currents of Eq.~\eqref{eq:current-mix}, 
\begin{equation}\label{eq:mu_ode}
    I(\mucol) = \pempty(\mucol, t) I_\mathrm{e}(\mucol) + \pfull(\mucol, t)I_\mathrm{f}(\mucol) = \frac{dq}{d\mucol} \cdot \frac{d\mucol}{dt}.
\end{equation}
Solving this differential equation together with the master equation~\eqref{eq:rate_equation} yields the dynamics of $\mucol$ presented in the main paper. The rates in Eq.~\eqref{eq:rate_equation} are here
\begin{equation} \label{eq:battery_rates}
    \ratetofull = \fcol(\edot), \ \ratetoempty = 1-\fcol(\edot),
\end{equation}
where $\fcol(\edot)$ is the electron distribution in the cold contact at the dot's energy level $\edot$. The rates are time-dependent due to the time-dependency in $\mucol$.

\subsection{$\tcharge$ and two-slope behavior}
For the feedback mechanism to be effective as intended, the quantum-dot level should be placed between the two filter heights, $\enone < \edot < \entwo$, since this enables the two-slope trajectory seen in Fig.~\ref{fig:supp_battery}(a) (and Fig.~\ref{paper-fig:ideal_result} of the main paper). Indeed, during significant ranges of time $\mucol(t)$ evolves linearly. This behavior can be understood by considering Eq.~\eqref{eq:mu_ode} further. Note that the differential equation is not analytically solvable, even if the probabilities are time-independent, meaning that a linear approximation is useful to estimate the trajectory analytically. 

First, we observe that $\mucol(t)$ evolves over an energy interval that is larger than $\kB\Tcol$ by an order of magnitude, meaning that $\pfull(t)$ is only affected by the thermal broadening in $\fcol(\en)$ for a small section of this interval; for most of the evolution, the quantum dot is either almost always empty or almost always occupied.  The magnitude of $\mucol(t)$ depends on the placement of $\enone$ and $\entwo$, along with the number of available electronic states per energy in the contact $qn_\mathrm{c}(\en)\equiv en_\mathrm{c}$, proportional to the capacitance $C$~\cite{Buttiker1993Sep,Pretre1996Sep,Gabelli2006Jul}. Higher filter heights allow for larger stopping voltages, while larger $qn_\mathrm{c}$, namely a larger capacitance, slows down the growth of $\mucol(t)$, increasing $\tcharge$. If these parameters have values that allow for $\mucol(t) > \kB \Tcol$, the charging trajectory will exhibit linear growth in certain time spans. 

These time spans are determined by the intervals in which $\pfull(t) \approx 0,1$  and by the filter heights. To see this, we consider the explicit integration of the current $I_\mathrm{e}$ (analogously for $I_\mathrm{f}$),
\begin{equation} \label{eq:integrated_current}
    I_\mathrm{e} = \frac{e}{h}\left(\kB\Tabs \ln\left[1+\exp\left(-\frac{\enone}{\kB\Tabs}\right)\right]-\kB\Tcol\ln\left[1+\exp\left(-\frac{\enone+\mucol}{\kB\Tcol}\right)\right]\right). 
\end{equation}
We are generally interested in a situation where $\Tabs \gg \Tcol$ which means the first term tends to dominate. Furthermore, for $\mucol \ll \enone$ the exponential in the second term is nearly zero, causing the term as a whole to almost vanish. This is the cause for the linear behavior: when the second term vanishes,  the left-hand side of Eq.~\eqref{eq:mu_ode} becomes independent of $\mucol$. Assuming that $\pfull = 0$ lets the ODE be solved by
\begin{equation} \label{eq:first_slope}
    \mucol(t) \approx \frac{1}{nh}\kB\Tabs \ln\left[1+\exp\left(-\frac{\enone}{\kB\Tabs}\right)\right]t, \quad \quad \mucol \ll \enone\ .
\end{equation}
For time spans when $\mucol \ll \enone, \entwo$ there is a linear dependence regardless of the value of $\pfull(t)$, since the second term in both $I_\mathrm{e}$ and $I_\mathrm{d}$ is suppressed. The value of the slope is then a linear combination of the first term of $I_\mathrm{e}$ and $I_\mathrm{d}$. But when $\mucol \gg \enone$ and $\mucol \ll \entwo$ the probability must be $\pfull \approx 1$, otherwise the second term in Eq.~\eqref{eq:integrated_current} for $I_\mathrm{e}$ becomes significant. If the linear approximation is valid, the second slope is determined by
\begin{equation} \label{eq:second_slope}
    \mucol(t) \approx \frac{1}{nh}\kB\Tabs \ln\left[1+\exp\left(-\frac{\entwo}{\kB\Tabs}\right)\right]t + K,  \quad \quad \edot \ll \mucol \ll \entwo
\end{equation}
where the constant $K$ depends on when the first and second curve intersect, which happens near $\mucol \approx \enone$. 

Having identified the two slopes, these are now used to estimate the charging time for each slope and in total. This lets us estimate the charging timescale, informing what values $\Gamma$ is allowed to take. We notice in Fig.~\ref{fig:supp_battery}(a) that $\mucol$ charges to a bit above $\enone$ for the first slope and a bit above $\entwo$ for the second slope. Since we are mainly interested in an order-of-magnitude estimation, we therefore choose for the linear approximation that the first slope stops at $\mucol = \enone$ and the second at $\mucol = \entwo$. Plugging $\mucol = \enone$ into Eq.~\eqref{eq:first_slope} gives
\begin{equation}
    \tau_\mathrm{charge, e} = \frac{\enone n h}{\kB\Tabs \ln\left[1+\exp\left(-\frac{\enone}{\kB\Tabs}\right)\right]}
\end{equation}
as the charging time associated with the first slope, when the quantum dot is mostly empty. 
From $\mucol  =\enone$ the battery continues to charge with Eq.~\eqref{eq:second_slope} until $\tau_\mathrm{charge}$ at $\mucol = \entwo$, assuming that the dot population immediately switches to $\pfull = 1$. Solving for the charging time in the same manner as for Eq.~\eqref{eq:second_slope} gives
\begin{equation} \label{eq:charging_time_K}
    \tau_\mathrm{charge} = \frac{nh(\entwo-K)}{\kB\Tabs\ln\left[1+\exp\left(-\frac{\entwo}{\kB\Tabs}\right)\right]}
\end{equation}
Next, $K$ in Eq.~\eqref{eq:second_slope} needs to be solved for by matching the two slopes, giving 
\begin{equation}
    K = \frac{\tau_\mathrm{charge, e}}{nh} \kB\Tabs\left(\ln\left[1+\exp\left(-\frac{\enone}{\kB\Tabs}\right)\right]-\ln\left[1+\exp\left(-\frac{\entwo}{\kB\Tabs}\right)\right]\right).
\end{equation}
Replacing this into Eq.~\eqref{eq:charging_time_K} yields
\begin{equation} \label{eq:charging_time}
    \tau_\mathrm{charge} = \frac{nh (\entwo-\enone)}{\kB\Tabs \ln\left[1+\exp\left(-\frac{\entwo}{\kB\Tabs}\right)\right]}  + \tau_\mathrm{charge, e} \equiv \tau_\mathrm{charge,f} + \tau_\mathrm{charge, e}
\end{equation}

From a purely technical perspective this is relevant, as we saw above: The ratio between $1/\Gamma$ and $\tcharge$ determines how well the averaged master equation, Eq.~\eqref{eq:rate_equation}, approximates the dynamics. In turn, the validity of the linear approximation depends on how quickly $\pfull(t)$ equilibrates to $\pfull(t)\approx 1$ when $\mucol \gg \edot$. 

\subsection{Charging dynamics outside linear approximation}
\begin{figure}[tbh]
    \centering
    \includegraphics[width=0.6\linewidth]{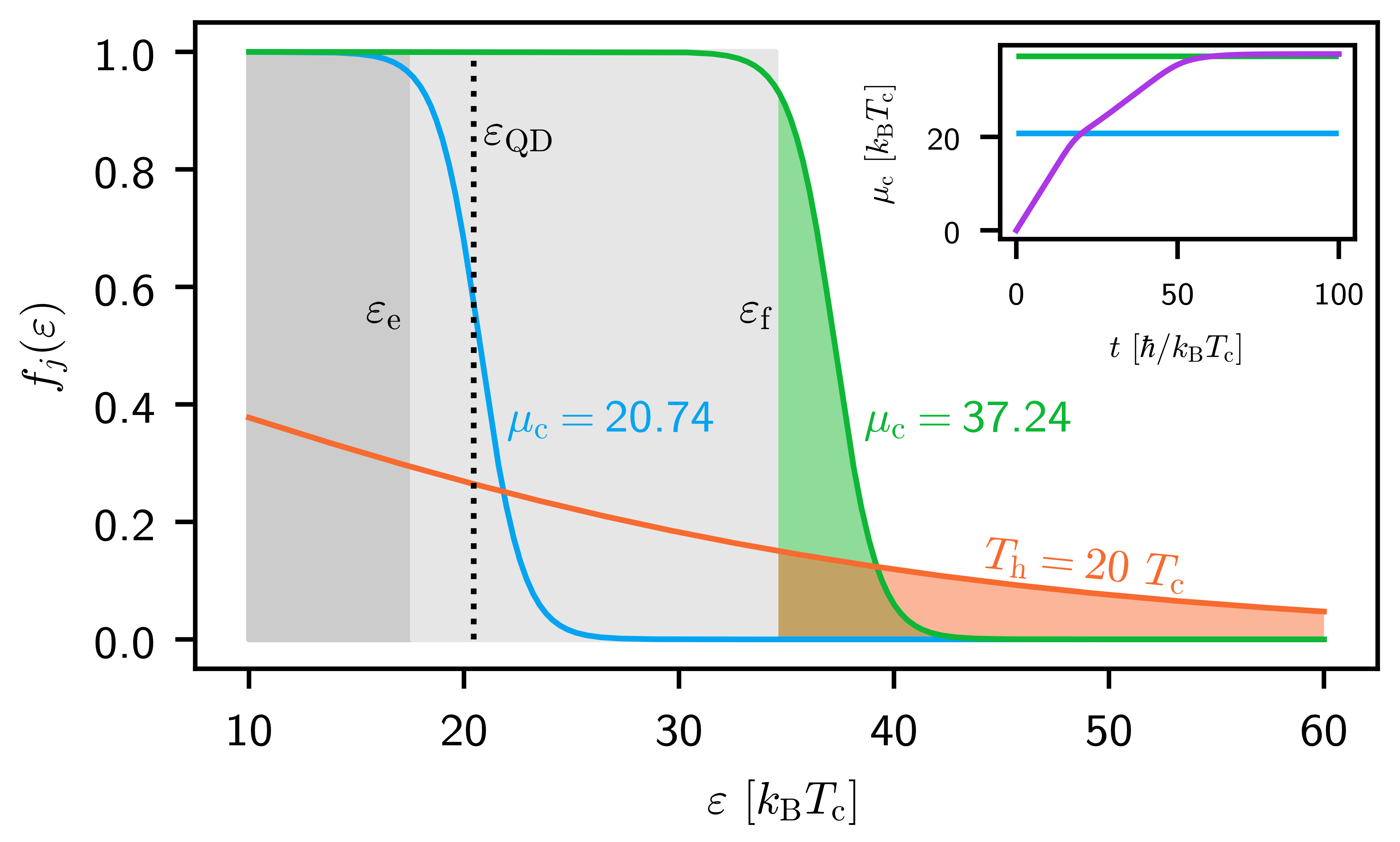}
    \caption{Electron distributions in the cold contact near the two ``bumps'' in the $\mucol(t)$ trajectory from Fig.~\ref{paper-fig:ideal_result}(a) in the paper, pictured again inside this figure.}
    \label{fig:battery_with_dist}
\end{figure}

In the time spans where a linear approximation cannot be made, $\mucol$ evolves according to the full ODE~\eqref{eq:mu_ode}, which does not have a closed analytical solution. Instead of analytical details, we discuss qualitatively what happens at the turning points with the help of Figure~\ref{fig:battery_with_dist}.
Figure~\ref{fig:battery_with_dist} shows the distribution function in the cold contact at two values of $\mucol$, each near turning points in the temporal trajectory as indicated in the inset, which is a copy of Fig.~\ref{paper-fig:ideal_result} in the main paper. The distribution functions are juxtaposed with the distribution in the hot contact (orange line). This serves to show the large temperature difference between the contacts. The first $\mucol$ reached, indicated by the blue line, is situated just above the quantum dot level $\edot$. Near this point, the quantum-dot population transitions from mostly empty to mostly full. Therefore, the trajectory moves from the first to the second slope as predominantly $\enfilter = \entwo$. Without the feedback the trajectory would reach the stopping voltage $\mucolstop$ associated with $\enone$. 

The second $\mucol$ demonstrated, represented by the green line, shows how $\mucolstop$ is reached. Recall here that the current is almost completely dominated by $I_\mathrm{f}$. As $\mucol$ increases the current necessarily decreases for a fixed $\fabs(\en)$, as is the case here. Eventually, the areas under the two distribution functions above $\entwo$ become equal and $\mucol$ can no longer increase as a consequence of the backflow. This constant $\mucol$ corresponds to the $\mucolstop$. 

\subsection{Approximating dot dynamics and estimating impact of $1/\Gamma$}

We are here interested in timescales that are much larger than $1/\Gamma$ yet much smaller than $\tcharge$ for the feedback setting introduced above. In this range, the master equation can to a good approximation be solved as
\begin{equation} \label{eq:battery_rate_solution}
    \pfull = \fcol(\edot) + \mathcal{C}_\mathrm{f} e^{-\Gamma t}.
\end{equation}
In the long-time limit of interest,$t \gg 1/\Gamma$, $\pfull$ hence converges to the quasi-steady-state probability 
\begin{equation} \label{eq:avg_pfull_battery}
    \avgpfull = \fcol(\edot)
\end{equation}
which still depends on time through $\mucol$. 

We are interested in a situation where the battery is charged from $\mucol = 0$. At this potential, typically the quantum dot is nearly guaranteed empty (especially for $\Tabs \gg \Tcol$) since $\edot > \enone$. Assuming then that the initial population is zero allows for the constant $\mathcal{C}_\mathrm{f}$ to be solved as
\begin{equation} \label{eq:init_constant}
    \mathcal{C}_\mathrm{f} = p_\mathrm{f}(t=0) -  \avgpfe = 0-\fcol(\edot,\mucol=0) \approx 0.
\end{equation}
This approximation, along with $1/\Gamma \ll \tau_\mathrm{charge}$ allowing for fast exponential decay relative to the charging time, justifies the replacement $\pfull \to \avgpfull$ in the differential equation \eqref{eq:mu_ode}. Then only Eq.~\eqref{eq:mu_ode} needs to be numerically solved instead of the combination with Eq.~\eqref{eq:rate_equation}. Nevertheless, in the main results, Fig.~\ref{paper-fig:ideal_result}(a) in the paper, the whole differential equation system is used in the numerical results by solving Eqs.~\eqref{eq:load_rate_eq} with rates~\eqref{eq:battery_rates} and Eq.~\eqref{eq:mu_ode} simultaneously. 
\begin{figure}[tbh]
    \centering
    \includegraphics[width=\linewidth]{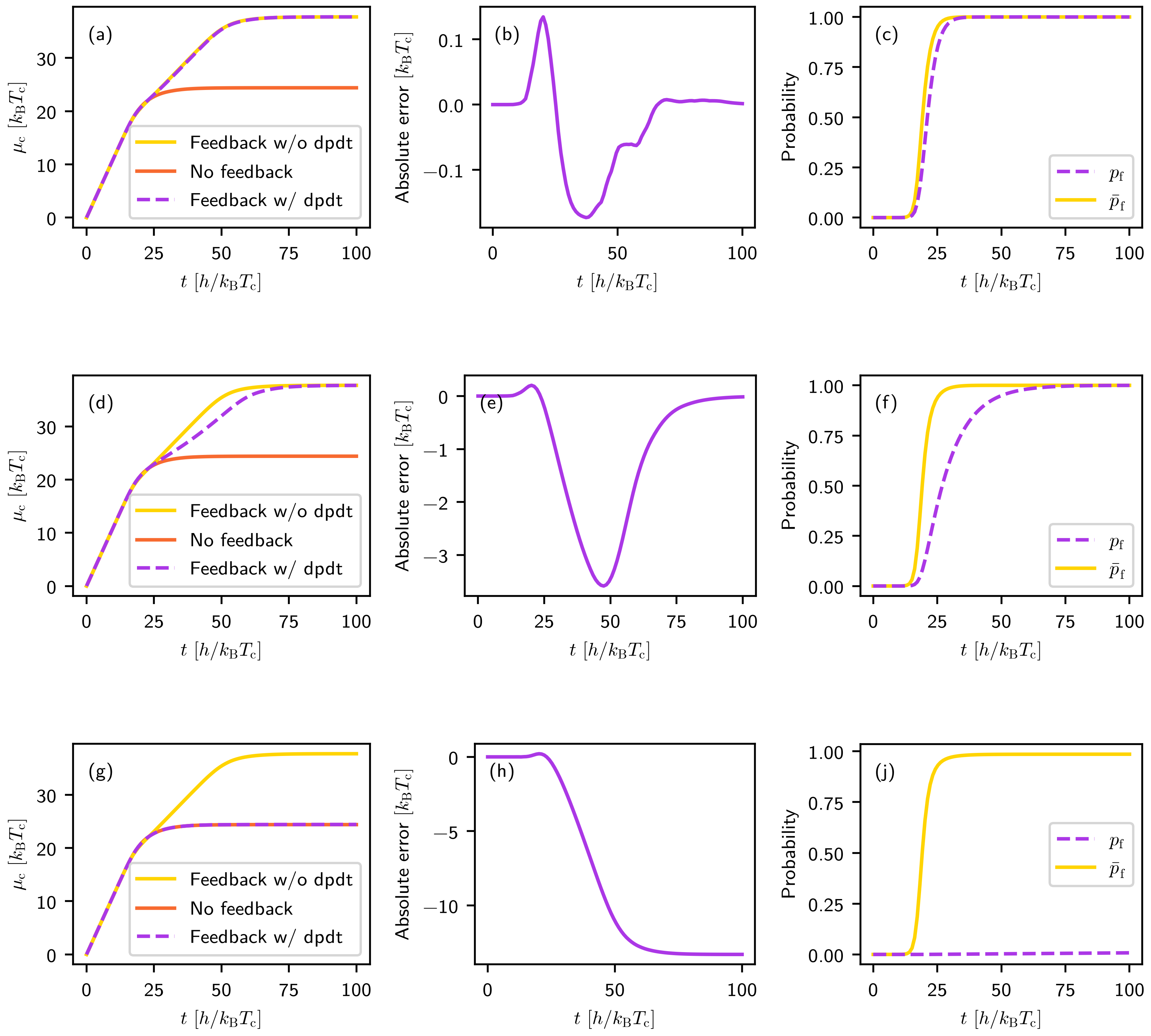}
    \caption{Trajectories for $\mucol$ with $1/\Gamma = 1$ over time for three cases: (red) no feedback, only transport with energy filter $\enone$; (purple) with feedback, $\mucol$ solved for assuming $\avgpfull$; (yellow, dotted) with feedback solving for $\mucol$ and $\pfull$ simultaneously. Parameters the same as in figure~\ref{paper-fig:ideal_result}(a) in the paper, except that the filter height for the nonfeedback curve is the same as $\enone$ for the feedback curve. Notably, the yellow and purple curves follow each other very closely, implying that the two methods for solving $\mucol$ are functionally equivalent.}
    \label{fig:supp_battery}
\end{figure}

Figure~\ref{fig:supp_battery}(a)-(c) shows that solving $\mucol$ either by solving both ODEs or by assuming $\avgpfull = \fcol(\edot)$ gives very similar results when $1/\Gamma = 2$. The battery setup has the same parameters as the feedback curve in Fig.~\ref{paper-fig:ideal_result}(a) of the main paper. Consequently, the charging time in Eq.~\eqref{eq:charging_time} is evaluated to $\tau_\mathrm{charge}/(en_\mathrm{c}) = 48.31\ \hbar/e$, which is close to the set charging time of $50\ \hbar/e$. Panel (a) demonstrates that the curves follow closely for either method with feedback. The dashed purple curve is the same as the solid purple curve in Fig.~\ref{paper-fig:ideal_result}(a) in the main paper. Panel (b) shows the absolute error between the two curves, with solving the rate equation minus without it. The error is around $10^{-1} \ \kB\Tcol$, which is two orders of magnitude smaller than the value of the curves, showing that the two methods are practically interchangeable within a small margin of error. Panel (c) compares the numerical solution to $\pfull$ (dashed purple) with $\avgpfull$ (yellow) as a function of $\mucol(t)$. The two curves follow closely to one another with only a small delay between the two. 

Figure~\ref{fig:supp_battery} should be carefully interpreted; the curves do not represent a single trajectory of $\mucol$, but rather the average trajectory over many charging cycles. As long as $\pfull(t)$ changes sufficiently fast, on a smaller timescale than the charging time, the expected trajectory can be physically interpreted as the approximate actual trajectory of $\mucol(t)$. 

The physicality of the trajectory begins to break down in Figure~\ref{fig:supp_battery}(d)-(f), which shows how the charging trajectory changes when $1/\Gamma = 10$, comparable to the charging time. Panel (d) shows that the trajectory for the complete ODE deviates from the trajectory using $\avgpfull$ after $\mucol > \enone$. Before this point $\mucol \ll \edot$, such that $\pfull$ is close to zero throughout the trajectory due to  initializing $\pfull(0) = 0$. Afterwards, $\mucol \gg \edot$ and the probability should equilibrate to one, but due to the low rate of jumps to the quantum dot, the trajectory takes longer to equilibrate, reaching the stopping voltage at a time closer to 70 than 50. Panel (e) shows that the error is markedly higher between the two curves under this operating condition. Panel (f) demonstrates that $\pfull(t)$ now has significant deviations from $\avgpfull(\mucol)$ showing that the approximate solution to the master equation in Eq.~\eqref{eq:battery_rate_solution} is becoming invalid. 

This breaks down fully in Figure~\ref{fig:supp_battery}(g)-(j), which shows the trajectories when $1/\Gamma = 10000$---much larger than the charging time. This is the regime used in the load setup. The dashed purple trajectory in panel (g) shows that $\mucol(t)$ is governed by $I_\mathrm{e}$ with $\pfull \approx 0$ for the entire interval, which is also evident in panel (c). Eventually, the average trajectory will reach the stopping voltage for the feedback curve, butit loses its physical meaning as the actual potential of the cold contact. Instead, the method for the load setup is more physically meaningful, where the physical $\mucol(t)$ is correlated with the quantum dot's trajectory.  

\section{Details on the rate equation for the load setup} 
The process for the load setup is as follows:
\begin{enumerate}
    \item An electron jumps in or out of the quantum dot, timescale $1/\Gamma$
    \item The energy filter level shifts up or down, instantaneously
    \item A new nonequilibrium steady-state is reached by current flowing and $\mucol$ adjusting, timescale $10/\Gamma-100/\Gamma$
    \item The process repeats when the quantum dot population changes again. 
\end{enumerate}

In the load setup we are interested in the opposite scenario from the battery setup with
\begin{equation}
    \tcharge \ll \frac{1}{\Gamma}.
\end{equation}
This means that the dot dynamics is much slower than the charging time of $\mucol$. Due to the timescale separation, $\mucol$ is considered to change instantly with the filter height as the dot gets emptied or filled. We are interested in the average power output which is established in the long-time limit, such that the charging trajectory has a small impact. It is instead relevant to find the timescale on which an average current and power are established and to ensure that it is much smaller than any temperature changes in time.

The current of interest is then the long-time limit of Eq.~\eqref{eq:current-mix},
\begin{equation} \label{eq:load_mixed_current}
    I = \avgpempty I_\mathrm{e} + \avgpfull I_\mathrm{f},
\end{equation}
with currents defined as in Eq.\eqref{eq:empty_full_currents}. In contrast to the battery setup, $\mucol$ takes on only two discrete values depending on if the dot is empty or full, $\mucolzero$ or $\mucolone$. The potentials are solved for by Ohm's law for each current $I_\mathrm{e/f}$ separately. The averaging of the total current is then encoded in the probabilities, so we wish to know the timescale of the steady-state probabilities $\avgpempty,\avgpfull$. 

The dynamics of the dot remain unchanged, Eq.~\eqref{eq:rate_equation}, and this informs the timescale for the average current and power. Due to the nature of the potential switching, the rates are now dependent on the electron distribution at two different potentials, 
\begin{equation} \label{eq:load_rate_eq}
    \frac{d\pfull}{dt}  = \ratetofull \pempty (t) - \ratetoempty \pfull (t) = \Gamma \fcol(\edot, \mucolzero ) \pempty (t) - \Gamma (1-\fcol(\edot, \mucolone )) \pfull (t)
\end{equation}

Using that $\pempty (t) = 1-\pfull (t)$ gives a straightforward solution to the differential equation, 
\begin{equation} \label{eq:load_rate_solution}
    \pfull (t) = \frac{\ratetofull }{\ratetofull  + \ratetoempty } + \mathcal{C}_\mathrm{f}e^{-(\ratetoempty +\ratetoempty )t},
\end{equation}
where $\mathcal{C}_\mathrm{f}$ is determined by the initial probability. From the solution for $\pfull (t)$ it is also straightforward to find the average occupation $\bar \pfull $ by taking $t\to \infty$. The exponential goes to zero, such that 
\begin{equation} \label{eq:average_prob}
    \avgpfull  = \frac{\ratetofull }{\ratetofull  + \ratetoempty }
\end{equation}
This is the probability that is used in Eq.~\eqref{eq:load_mixed_current} to find the mixed current and furthermore the mixed power 
\begin{equation} \label{eq:power_mixture}
    P = \avgpfull \frac{\mucolone^{2}}{eR} + \avgpempty \frac{\mucolzero^{2}}{eR}
\end{equation}

\begin{figure}[tbh]
    \centering
    \includegraphics[width=0.6\linewidth]{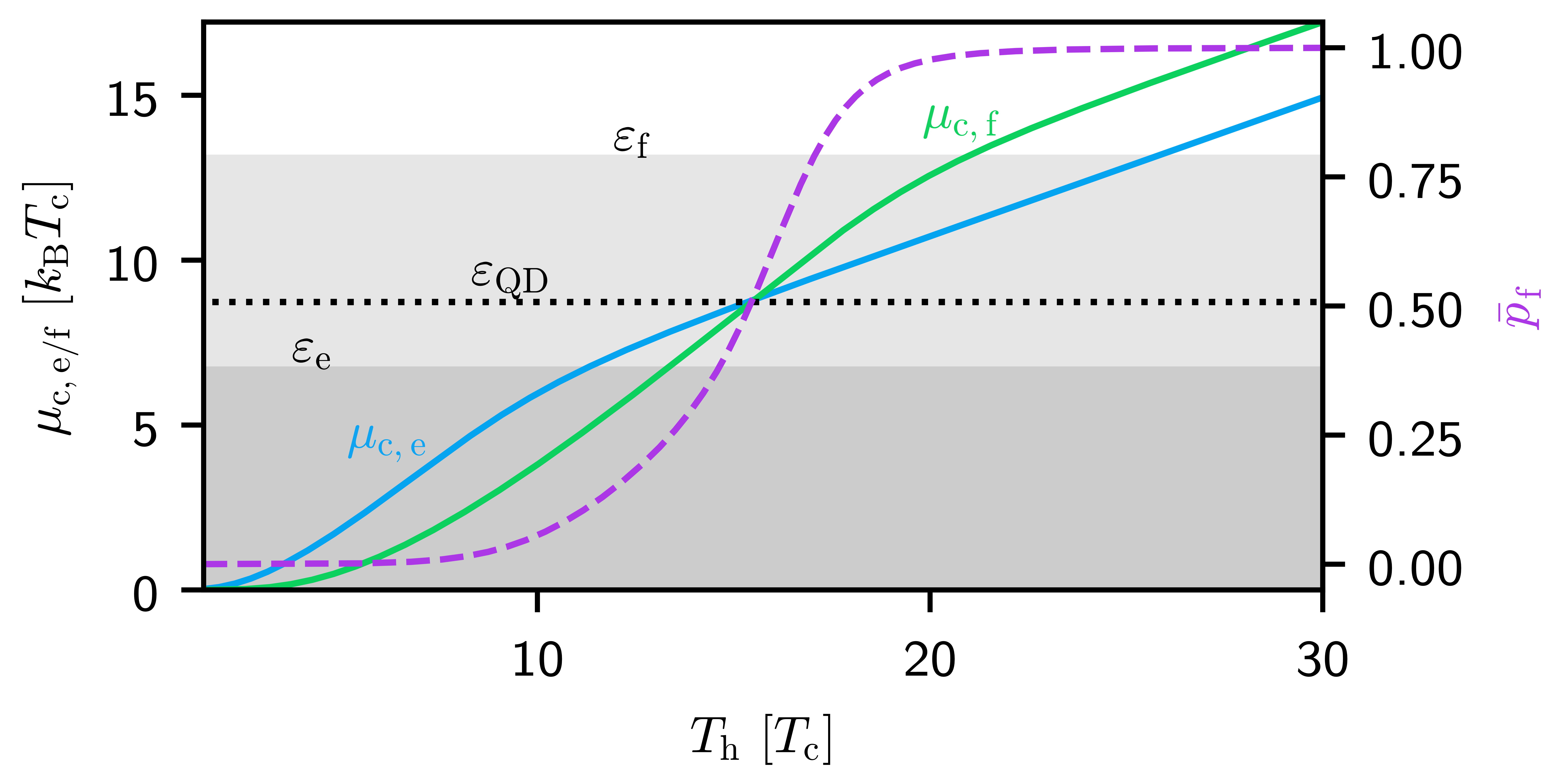}
    
    \caption{Plots of $\mucolzero$ and $\mucolone$ for the optimized load setup with parameters as in Fig.~\ref{paper-fig:temp_dist_results}(b) in the main paper, and Fig.~\ref{fig:full_bimodal_result}(d), pictured along with $\avgpfull$ on the right y-axis. }
    \label{fig:mu_for_power}
\end{figure}

The potentials $\mucolzero$ and $\mucolone$ depend highly on $\Tabs$, and consequently so does $\avgpfull$ and dependent quantities. In the main paper we present the power output over a range of temperatures, Fig.~\ref{paper-fig:temp_dist_results}, which is furthermore repeated in Fig.~\ref{fig:full_bimodal_result}. To understand how the power curve is formed, we consider Figure~\ref{fig:mu_for_power}. It shows how the $P_\mathrm{opt}$ in Fig.~\ref{paper-fig:temp_dist_results}(b) saturates the bound at two points. Up to the first point $\avgpfull \approx 0$ where $\mucolone(\Tabs), \mucolzero(\Tabs) \lesssim \edot$ such that only $\mucolzero(\Tabs)$ contributes to the power. Then follows a transition as $\avgpfull \approx 0 \to \avgpfull \approx 1$ with $\mucolone(\Tabs), \mucolzero(\Tabs) \lesssim \edot$ such that only $\mucolone$ contributes, which satureates the bound at the second point. The power curve ``moves'' from the curve $\mucolzero^2/eR$ to $\mucolone^2/eR$ as the dot goes from almost guaranteed empty to almost guaranteed full. 

\subsection{Convergence time}
While only the average population probability is needed for the computations, it is prudent to establish at what timescale the dot population stabilizes. Intuitively, if the potentials $\mucolzero$ and $\mucolone$ are far apart, and the quantum dot level lies in between the two, then the dot is unlikely to fill when it is empty and unlikely to empty when it is full. Both the rates $\ratetoempty $ and $\ratetofull $ are then small, necessitating that $t$ is large for the population probability to converge to the steady-state, $\pfull(t) \to \avgpfull$. Calling the time it takes for the probabilities to converge to their steady-states the convergence time $\tconv$, it is limited by the timescale on which the incoming temperature changes, which depends on the specific environment of the device. Although such an environment is not specified in the main text, it is useful to determine how that could be taken into account. The time $\tconv$ also informs when the timescale for the average current Eq.~\eqref{eq:load_mixed_current}. 

First, the convergence time $\tconv$ needs to be concretely defined. To this end, the convergence time for $\pfull(t)$ specifically is introduced, $\tau_\mathrm{conv,f}$. To completely reach $\bar \pfull $, the time needs to be $t \to \infty $, but in practice there is some $\tau_\mathrm{conv,f}$ when $\pfull (t)$ approaches closely enough to $\bar \pfull $ such that the device description is accurate enough. The accuracy one would like to achieve can be quantified by a tolerance $\eta$ as,
\begin{equation}
    |\bar \pfull  - \pfull (\tau_\mathrm{conv,f})| = \eta.
\end{equation}
With Eq.~\eqref{eq:load_rate_solution} $\tau_\mathrm{conv,f}$ is solved for as
\begin{equation} \label{eq:conv_time_f}
    \tau_\mathrm{conv,f}  =  -\frac{1}{\ratetoempty +\ratetoempty }\ln\left(\frac{\eta}{|\mathcal{C}_\mathrm{f}|}\right).
\end{equation}
The convergence time $\tau_\mathrm{conv,e}$ for $\pempty$ is found in the same manner. From Eq.~\eqref{eq:init_constant} one finds that $\mathcal{C}_\mathrm{e} = -\mathcal{C}_\mathrm{f}$, meaning that the two convergence times are the same and
\begin{equation} \label{eq:conv_time_full}
    \tconv \equiv \tau_\mathrm{conv,e} = \tau_\mathrm{conv,f}. 
\end{equation}
The constant $\mathcal{C}_\mathrm{f}$ is a measure of how far the initial dot population is from the steady-state probability, which becomes apparent in that the convergence time is positive only for $\eta < |\mathcal{C}_\mathrm{f}|$---the probability is already converged with a lower tolerance than $\eta$. We will typically assume the quantum dot starts empty, $\pfull(t= 0) = 0$. 

With the convergence time defined, we can estimate the timescale for the averaging of $I$ by finding the maximum possible convergence time for a circuit with given filter heights but variable $\edot$. In the next section we will also investigate how the timescale depends on the filter heights. 

\begin{figure}
    \centering
    \includegraphics[width=\linewidth]{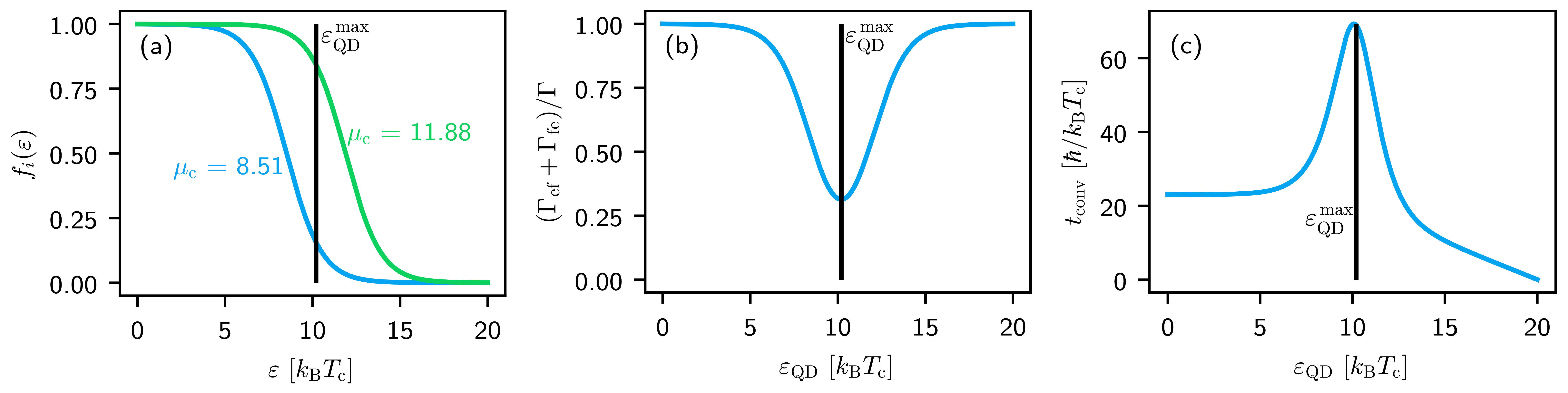}
    \caption{Example plots for the convergence time for the load setup, with  $\Tabs = 20 \ \Tcol $, $\enone = 0 \ \kB\Tcol$, $\entwo = 10 \ \kB\Tcol$, $\Gamma = 1/2$. (a) Electron distributions in the cold contact for $\mucolzero$ in blue and $\mucolone$ in green, with $\edot^\mathrm{max}$ marking the energy in the middle of the two potentials, leading to $\avgpfull = 0.5$. (b) The sum of the rates in Eq.~\eqref{eq:load_rate_eq} normalized with $\Gamma$ depending on $\edot$. The minimum is reached at $\edot^\mathrm{max}$. (c) The convergence time for $\pfull(t)$ as defined in Eq.~\eqref{eq:conv_time_full} with $\eta = 10^{-5}$ and $p_\mathrm{f, init} = 0$. The maximum is reached near $\edot^\mathrm{max}$.}.  
    \label{fig:conv_dist}
\end{figure}

To illustrate how the estimate is made, an example system is employed with $\Tabs = 20\Tcol$. The filter heights are set to $\enone = 0 \ \kB\Tcol$ and $\entwo = 10 \ \kB\Tcol$. Solving for the potential at each filter height gives the two distributions in figure~\ref{fig:conv_dist}(a).

To find the upper bound of the convergence time we look at each of the two factors in Eq.~\eqref{eq:conv_time_f} separately. First, the logarithm contributes the most when $|\mathcal{C}_\mathrm{f}| = 1$, meaning that the initial and average probabilities are opposite.

Next, we consider how the $\edot$ influences the other factor, $1/(\ratetoempty + \ratetofull)$. 
The sum of the rates can go to zero, causing the convergence time to diverge---if the dot level $\edot = \edot^\mathrm{max}\equiv(\mucolone + \mucolzero)/2$, as pictured in Fig.~\ref{fig:conv_dist}(a). At this point, $\mucolzero \ll \edot$ meaning that the dot tends to remain empty once it becomes empty, and $\mucolone \gg \edot$ so the dot remains full once it has been filled. Thus the dot state can get ``stuck'', causing long convergence time since the steady-state probability is $\avgpfull = 1/2$, requiring that the dot switches state many times. Panel (b) shows the sum of the rates depending on $\edot$, clearly demonstrating how the sum approaches zero for $\edot= \edot^\mathrm{max}$ (vertical black line). 

This dot level placement is the point where $\ratetoempty  = \ratetofull $, giving $\bar \pfull  = 1/2$. Both $p_\mathrm{f,\mathrm{init}} = 0$ and $p_\mathrm{f,\mathrm{init}} = 1$ then give $\mathcal{C}_\mathrm{f} = 1/2$. Using this constant, figure~\ref{fig:conv_dist}(c) shows the convergence time $\tconv$ as a function of $\edot$ for the example system. Using $\edot = \edot^\mathrm{max} \equiv (\mucolone + \mucolzero)/2$ does not exactly align with the maximum, since $\ln(\eta/|\mathcal{C}_\mathrm{f}|)$ slightly skews the maximum. To maximize the estimation, we use the rates for $\edot = \edot^\mathrm{max}$, but take $C_\mathrm{f} = 1$. That gives the largest possible value of the convergence time, 

\begin{equation} \label{eq:max_conv_load}
    \tconv^\mathrm{max} =-\frac{1}{\Gamma}\frac{1}{2\fcolempty((\mucolone + \mucolzero)/2)} \ln(\eta)
\end{equation}

\subsection{Convergence depending on $U$}
For the scenario where $\edot = (\mucolone + \mucolzero)/2$, $\edot > \enone$,  we would now like to estimate the impact of the capacative coupling $U = e^2/C = \entwo-\enone$ by expressing Eq.~\eqref{eq:max_conv_load} as a function of $U$ and finding an upper bound on $U$; a larger capacative coupling gives a larger difference $\mucolone - \mucolzero$, in turn giving lower rates on the dot dynamics. The goal is to ensure that the maximum convergence time in Eq.~\eqref{eq:max_conv_load} is lower than the timescale for $\Tabs$, or some other relevant timescale if desired, and to understand how the QPC's properties affect the timescale for averaging $I$.  

First, the potential values are estimated for the largest difference $\mucolone-\mucolzero$. From Ohm's law, $\mu = eRI$, the equations for the potentials are
\begin{subequations}\label{eq:load_potentials}
    \begin{eqnarray}
    \mucolzero = \frac{e^2}{h}R\kB\Tabs\ln\left(1+\exp\left(\frac{-\enone}{\kB\Tabs}\right)\right) - \frac{e^2}{h}R\kB\Tcol\ln\left(1+\exp\left(\frac{-\enone+\mucolzero}{\kB\Tcol}\right)\right), \label{eq:load_potentials_a}\\
        \mucolone = \frac{e^2}{h}R\kB\Tabs\ln\left(1+\exp\left(\frac{-\entwo}{\kB\Tabs}\right)\right) - \frac{e^2}{h}R\kB\Tcol\ln\left(1+\exp\left(\frac{-\entwo+\mucolone}{\kB\Tcol}\right)\right).       
    \end{eqnarray}
\end{subequations}
The filter height $\enone$ is the optimal filter height for some temperature $\Tabszero$ and equivalently for $\entwo$ and $\Tabsone$. We want to find an upper bound on $U$ such that the probability $\pfull$ converges for any $\enone < \edot < \entwo$ and for any temperature $\Tabs \gg \Tcol$. Concentrating on Eq.~\eqref{eq:load_potentials_a}, the exponentials can be approximated for certain parameter values: for the first term, 
\begin{equation}
    \frac{e^2}{h}R\kB\Tabs\ln\left(1+\exp\left(\frac{-\enone}{\kB\Tabs}\right)\right) \approx \frac{e^2}{h}R\kB\Tabs \left(\ln(2) - \frac{\enone}{2\kB\Tabs}\right), \kB\Tabs \gg \enone,
\end{equation}
and for the second term,
\begin{equation}
    \frac{e^2}{h}R\kB\Tcol\ln\left(1+\exp\left(\frac{-\enone+\mucolzero}{\kB\Tcol}\right)\right) \approx \frac{e^2}{h}R\kB\Tcol\ln\left(\exp\left(\frac{-\enone+\mucolzero}{\kB\Tcol}\right)\right) = \frac{e^2}{h}R\kB\Tcol\left(\frac{-\enone+\mucolzero}{\kB\Tcol}\right), \mucolzero \gg \enone.
\end{equation}
Putting both the terms together again gives
\begin{equation}
    \mucolzero \approx \frac{Re^2/h}{1+Re^2/h} \left(\kB\Tabs \ln(2) + \frac{\enone}{2} \right).
\end{equation}
In the same manner, $\mucolone$ is approximated as
\begin{equation}
    \mucolone \approx \frac{Re^2/h}{1+Re^2/h} \left(\kB\Tabs \ln(2) + \frac{\entwo}{2} \right).
\end{equation}
These approximations are valid for all $\Tabs \gg \Tabsone$, giving the maximum potential difference
\begin{equation} \label{eq:mu_diff_approx}
    \mucolone-\mucolzero \approx \frac{Re^2/h}{1+Re^2/h}\frac{\entwo-\enone}{2} \equiv \frac{\varphi U}{2},
\end{equation}
in other words the potential difference is constant with increasing $\Tabs$. 
Since the linear asymptots are reached by growing values in exponentials, the linearity is reached quickly with growing $\Tabs$. It is therefore reasonable that even at $\Tabs>\Tabsone$ we can approximate that largest $\mucolone-\mucolzero$ for $\edot$ within $\enone$ and $\entwo$ to $\mucolone-\mucolzero \approx\frac{\entwo-\enone}{2} = U/2$. If we place $\edot$ in between these two potentials we should get the largest possible $\tconv$ for the setup. For a continuous range of $\Tabs$, $\edot$ will be placed between $\mucolzero(\Tabs)$ and $\mucolone(\Tabs)$ for some $\Tabs$. If this is true for some $\Tabs > \Tabsone$, then the convergence time is maximized. 

\begin{figure}[tbh]
    \centering
    \includegraphics[width=0.5\linewidth]{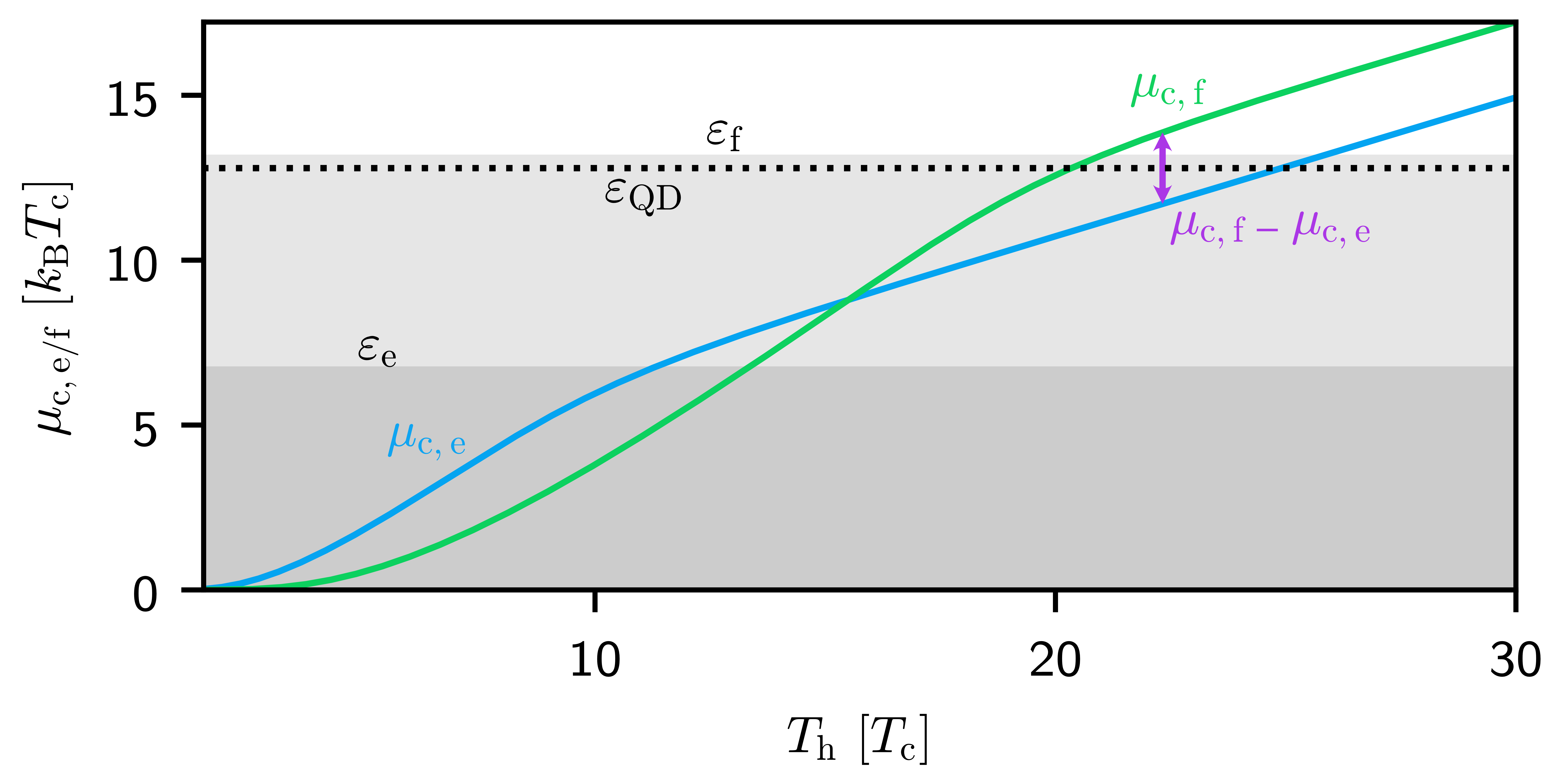}
    \caption{Potentials at empty or full quantum dot for the load setup. The quantum dot level $\edot$ is placed to maximize convergence time, which here happens at $\Tabs = 22.33 \ \Tcol$. Parameters, except $\edot$, the same as for Fig.\ref{paper-fig:ideal_result}(b) in the main paper. }
    \label{fig:qd_place}
\end{figure}

Figure~\ref{fig:qd_place} helps explain the arguments; it depicts $\mucolzero,\mucolone$ as functions of $\Tabs$ and shows that $\mucolone$ crosses $\mucolzero$ between the filter heights. Thus under the constraint that it's possible for $\edot = (\mucolone+\mucolzero)/2$ while $\enone<\edot<\entwo$, the largest difference $\mucolone(\Tabs)-\mucolzero(\Tabs)$ occurs for $\Tabs> 20$; here we can see that both potentials approach their linear asymptots. We choose $\edot$ so it is placed between the potentials at $\Tabs = 22.33 \ \Tcol$. The purple arrow indicates the potential difference at this temperature, which is evaluated to $\mucolone-\mucolzero = 2.17 \ \kB\Tcol$. To compare, $\varphi (\entwo-\enone)/2 = 1.97 \ \kB\Tcol$---slightly lower but a fairly close approximation. 

Having found an approximation of the maximum potential difference, Eq.~\eqref{eq:mu_diff_approx}, we plug it into the maximum convergence time in Eq.~\eqref{eq:max_conv_load} and get 

\begin{equation} \label{eq:conv_with_U}
    \tconv^\mathrm{max} \approx -\frac{1}{\Gamma} \frac{1}{2f_\mathrm{c,0}(\varphi U/4)}\ln(\eta).
\end{equation}
Here
\begin{equation}
    f_\mathrm{c,0}(\varphi U/4) = \frac{1}{1+\exp(\varphi U/(4\kB\Tcol))}.
\end{equation}
With this expression, it is now possible to estimate the timescale needed for time-averaging the current by only system parameters and the desired tolerance. These are then the parameters that enter in Table I in the main paper. Setting $\eta = 10^{-5}$ as a reasonable tolerance and the filter heights as in Fig.~\ref{paper-fig:ideal_result}(a) in the main paper gives $\tconv^\mathrm{max} = 21.19/\Gamma$. This motivates the timescale range in Table I in the main paper.

Solving for $U$ yields
\begin{equation} \label{eq:max_U}
    U^\mathrm{max} = \frac{4\kB\Tcol}{\varphi} \ln\left(-\frac{2\Gamma \tconv^\mathrm{max}}{\ln \eta}-1\right)
\end{equation}
as the upper bound on the coupling $U$ for a given convergence time, determined for instance by a desired timescale for the average current. To emphasize, the dot level does not \textit{need} to be placed such that the convergence time is maximized; $U$ could be larger if $\edot$ is placed close to $\enone$, for instance. The bound provides a guideline to achieve tolerable convergence times without constraints on $\edot$. 

As an example, let us calculate $U^\mathrm{max}$ for reasonable parameters with respect to the main text. In natural units normalized to $\Tcol$, $\hbar = 1, \kB \Tcol = 1$. We set $\Gamma = 1/2$ for simplicity; then the maximum convergence time could be a few orders of magnitude higher, say $\tconv^\mathrm{max} = 100$. The average current in Eq.~\eqref{eq:load_mixed_current} should be reliable, so we set $\eta = 10^{-5}$ as a reasonable tolerance on the probabilities. Plugging these numerical values into Eq.~\eqref{eq:max_U} gives
\begin{equation}
    U^\mathrm{max} = 13.28\ \kB\Tcol
\end{equation}
The results in the paper have couplings lower than this, which means that those feedback systems should have convergence times within a reasonable timescale. We stress, however, that what timescale is considered reasonable depends on the environmental context of the device, which is not addressed in detail here. 

Another note is that $|\mucolone - \mucolzero|$ could be larger than the difference between the two asymptots if $\enone > U$. Then it could be that $\mucolzero - \mucolone \approx \mucolzero$ near $\Tabs \approx \Tabszero$, which could be larger than Eq.~\eqref{eq:mu_diff_approx}. However, for a practical feedback device, the dot level would never be placed below $\enone$, because the feedback would never kick in. The bound on $U$ is therefore not constrained by this potential difference in practicality. 

\subsection{Realistic $U$}
Let us finally evaluate the capacative coupling $U$ in physical units in order to compare with experimental data. In our calculations (Fig.~\ref{fig:add_results}), $U \sim 10 \ \kB\Tcol$. For an order of magnitude estimation, let $\Tcol \approx 100$ mK. Then
\begin{equation}
    U \approx 8.62 \times 10^{-5} \ \mathrm{eV}.
\end{equation}
To make a simple comparison with existing literature, we set the temperature equal to zero on both sides. Then the difference in currents for an empty and full quantum dot is
\begin{equation}
    |I_\mathrm{f} - I_\mathrm{e}| = \frac{e}{h}U.
\end{equation}
We can then evaluate 
\begin{equation}
    \frac{e}{h} = 0.39 \times 10^{-4} \ \mathrm{A/eV}
\end{equation}
which gives 
\begin{equation}
    \frac{e}{h}U = 3 \ \mathrm{nA}
\end{equation}
This is very close to the observed current difference in Ref.~\cite{Gustavsson2006}, where the temperature is equal on both sides and much smaller than the electrochemical potentials. In general, the QPC current seems to range between pA and nA in experiments \cite{Lu2003May,Schleser2007,Fujisawa2006}. For the normal distribution in particular, $U$ is arguably closer to $1 \ \kB \Tcol$, approaching $I\sim 0.1 \ \mathrm{nA}$ as in Refs.~\cite{Lu2003May,Fujisawa2006}. We therefore conclude that the couplings $U$ that we find are reasonably sized for a realistic setting. 

\section{Other temperature distributions}
\begin{figure}[h!]
    \centering
    \subfloat[Results for the bimodal temperature distribution, constructed by normal distributions centered at $ 10 \Tcol$ and $ 20 \Tcol$ with standard deviation $\sigma =  \Tcol$. ]{\includegraphics[width=0.45\linewidth]{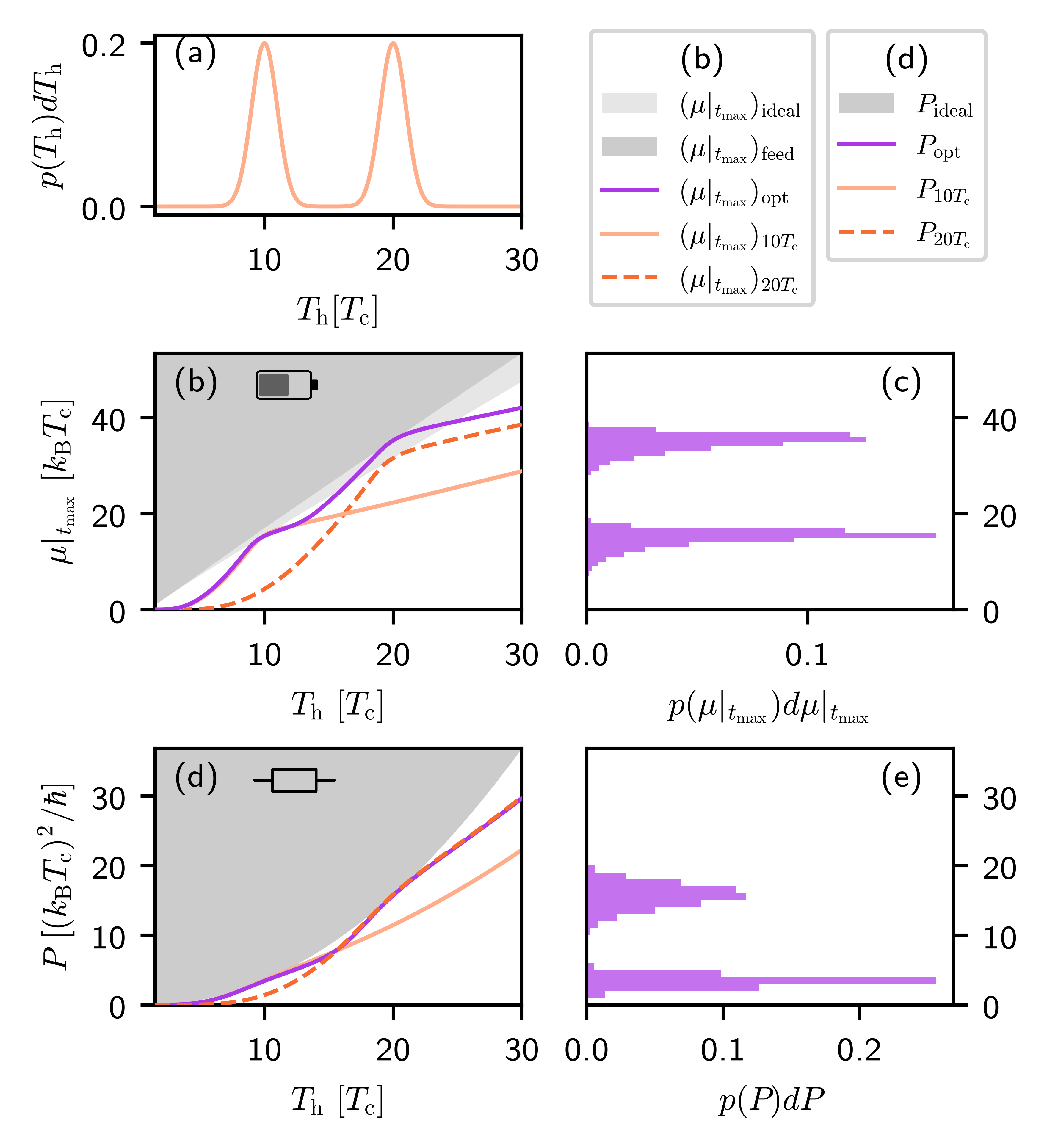}    \label{fig:full_bimodal_result}}    \subfloat[Results for the normal temperature distribution centered at $\Tabs = 20\Tcol$ with standard deviation $\sigma = 3\Tcol$]
    {\includegraphics[width=0.45\linewidth]{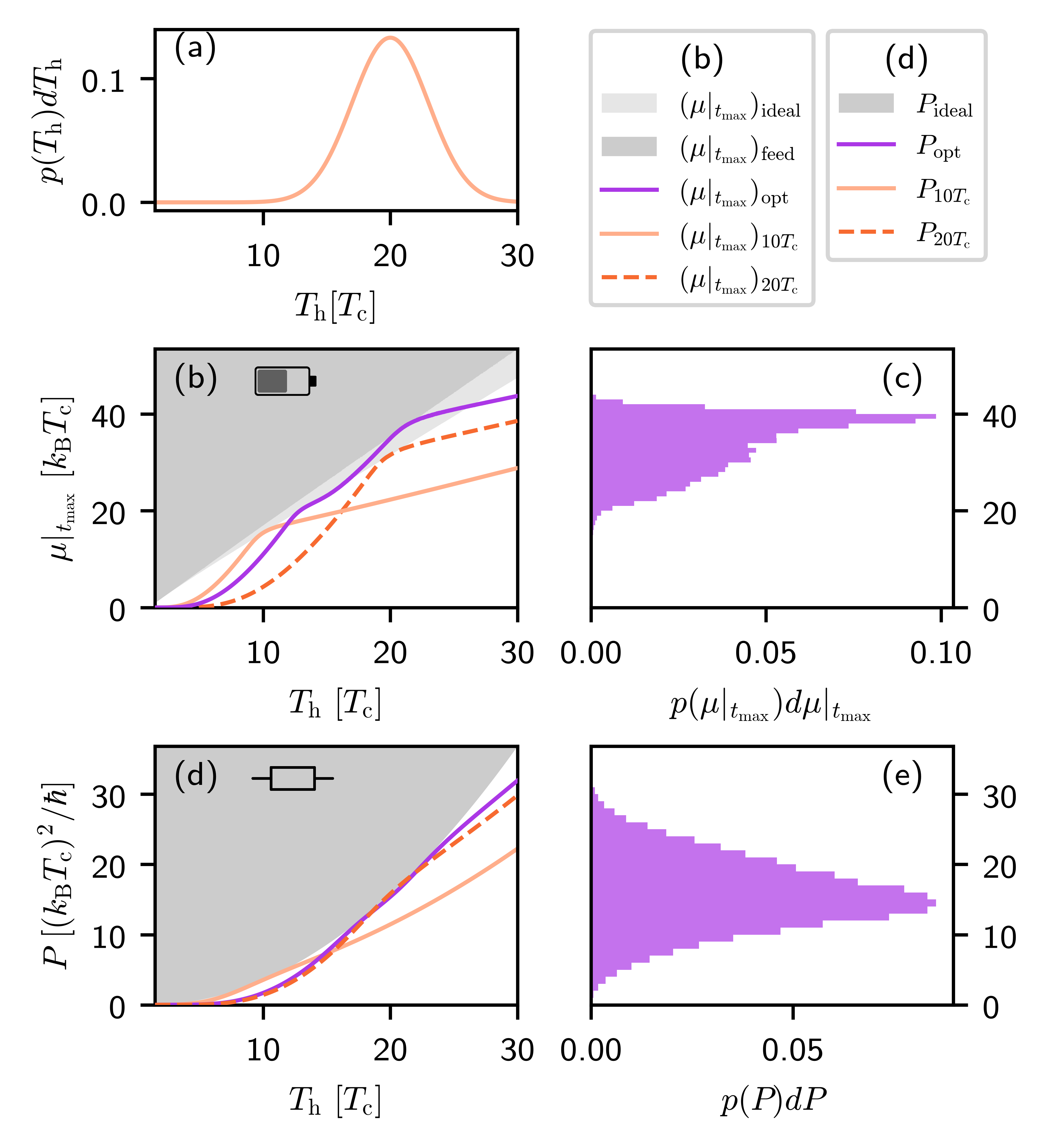}\label{fig:normal_results}}
    \hfill 
    \subfloat[Results for the uniform temperature distribution defined between $\Tabs = 1.5\Tcol$ and $\Tabs = 30\Tcol$.]{\includegraphics[width=0.45\linewidth]{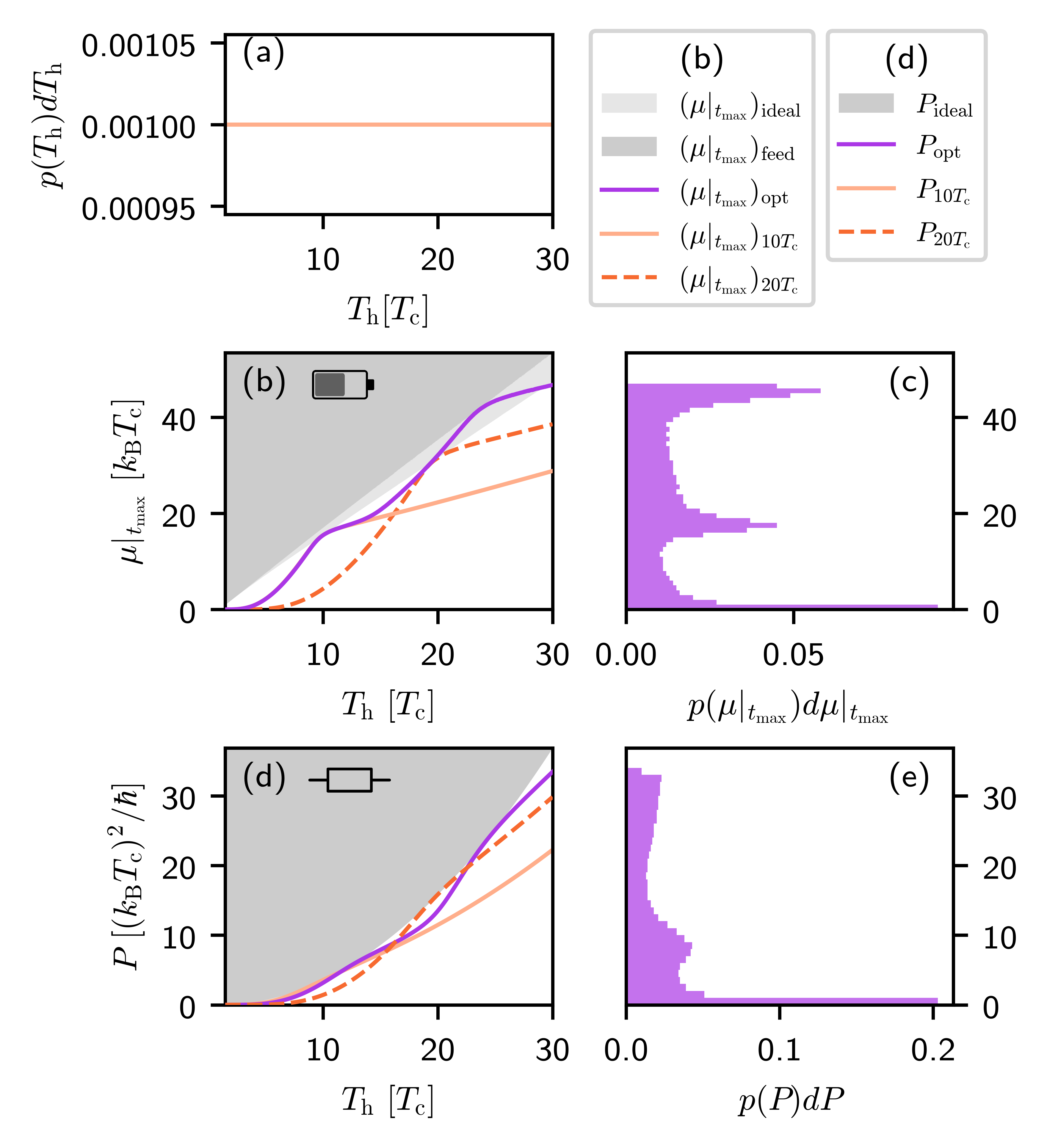}\label{fig:uniform_results}}
    \hfill
    \caption{Results for three temperature distributions: (i) Normal, (ii) Uniform, (iii) Bimodal. In each subfigure the panels depict: (a) The probability distribution. (b) Charging potentials $\mu|_{t_\mathrm{max}}$ at $\tcharge/(en_\mathrm{c}) = 50 \ \hbar/e$. The legend (b) in the upper right corner describes the curves. For (b) $(\mu|_{t_\mathrm{max}})_\mathrm{feed}$ is the highest possible $\mu$ for each $\Tabs$ \textit{with} optimal feedback. The other subscripts are equivalent to those in (d). Parameters for optimized feedback curve (purple) for each temperature distribution are found in Table~\ref{tab:average_values}. See Fig.~\ref{paper-fig:ideal_result}(a) in the main paper for other parameters. For all figures $\Gamma = 1/2 \ \kB \Tcol / \hbar$. (c) Probability distribution for $(\mu|_{t_\mathrm{max}})_\mathrm{opt}$, see ~Eq.~\eqref{paper-eq:average_power} in the main paper. (d) Produced power in the load setup: $P_\mathrm{ideal}$ is achieved when the barrier height is optimal at each temperature; $P_{10\Tcol}$ and $P_{20\Tcol}$ are the powers when the barrier height is optimal for $\Tabs = 10 \ \Tcol$ and $\Tabs = 20 \ \Tcol$ respectively, see Fig.~\ref{paper-fig:ideal_result}(b) in the main paper for their parameters; $P_\mathrm{opt}$ is the power for feedback with filter heights optimized for maximum average power given the temperature distribution of (a), see parameters in Table~\ref{tab:average_values} for each distribution. For all curves, $R = 10 \ e^2/\hbar$(e) Analogous to (d) for $P_\mathrm{opt}$.
    \label{fig:add_results}} 
\end{figure}

In the paper, a bimodal distribution is used as the probability distribution for $\Tcol$, see Fig.~\ref{paper-fig:temp_dist_results}. Here, we present results for two additional temperature distributions: a normal one and a uniform one.  For all three distributions, we also present results for battery setups subject to temperature distributions in addition to results for the load setup described in the main paper. Here, we analyze a situation where the battery is allowed to charge for a set time before it is disconnected from the system. The temperature $\Tabs$ is constant throughout the charging process, representing a situation where the exact temperature is unknown, such that the device is initialized at some $\Tabs$ according to a probability distribution. We choose the allowed charging time $t_\mathrm{max}/(en_\mathrm{c}) = 50 \ \hbar/e$ and optimize the feedback mechanism to maximize the average potential at this time, $\overline{(\mu|_{t_\mathrm{max}})}_{\mathrm{opt}}$---this is the potential one would expect on average for repeated initializations of the device. Note that we always set $\mucol(\Tabs)|_{t=0} = 0$ for all temperatures $\Tabs$, meaning that there is no dependence on previous temperatures. 

Figure ~\ref{fig:add_results} shows curves for all the results. In Table~\ref{tab:average_values}, averaged quantities are presented. We are interested in the average power or potential being as high as possible under the assumption that the temperature obeys a probability distribution. Before discussing the individual distributions, we see from the table that all average quantities with feedback and optimized parameters are larger than those for no-feedback devices optimized at $\Tabs = 10\Tcol$ or $\Tabs = 20\Tcol$. This indicates that the feedback mechanism is useful for a variety temperature distributions and for both setups. 

Concentrating first on the bimodal distribution, Fig.~\ref{fig:full_bimodal_result} repeats the results in Fig.~\ref{paper-fig:temp_dist_results} of the main paper, with the addition of the battery setup in panels (b) and (c). In panel (b), we see that there are two bounds on the largest possible $\mucol|_{t_\mathrm{max}}$: one tighter if there is no feedback (lighter grey) and one looser if there is feedback ideally chosen for each temperature (darker grey). Compare with Fig.~\ref{paper-fig:ideal_result}(a) in the main paper where values of $\mucol$ at $t/(en_\mathrm{c}) = 50 \ \hbar/e$ give the nonfeedback and feedback bounds at $\Tabs = 20 \ \Tcol$, since the curves are optimized for that temperature. The purple curve in Fig.~\ref{fig:full_bimodal_result}(b) shows how $(\mu|_{t_\mathrm{max}})_{\mathrm{opt}}$ varies with $\Tabs$ in the presence of feedback optimized for the bimodal temperature distribution. The feedback curve has two ``bumps'', where the first saturates the nonfeedback bound and the other saturates the bound with feedback. This is because the optimal value of $\edot$ is found in between the two  barrier heights $\enone,\entwo$, meaning that for low temperatures where $\mucol \lesssim \enone$ the feedback mechanism never kicks in. Similar to the load setup, the potential curve follows the two no-feedback curves, saturating a bound near $\Tabs = 10\Tcol$  and $\Tabs = 20\Tcol$. The difference for the potential is that the purple line saturates the higher bound for feedback-enabled charging processes at $\Tabs = 20\Tcol$.  
Fig.~\ref{fig:full_bimodal_result}(c) shows the probability distribution for $(\mu|_{t_\mathrm{max}})_{\mathrm{opt}}$ where the two peaks clearly pick potential values near $\Tabs = \ 10\Tcol$ and $\Tabs = 20 \Tcol$.

Next, Fig.~\ref{fig:normal_results} shows that the equivalent results to Fig.~\ref{fig:full_bimodal_result} but for a normal distribution centered at $\Tabs = 20\Tcol$. Both the power and potential curves for setups with feedback are shifted from Fig.~\ref{fig:full_bimodal_result}, since the parameters, see Table~\ref{tab:average_values}, are optimized for the normal temperature distribution. Here, it is beneficial for the average quantities to saturate the bound near $\Tabs = 20 \Tcol$, but also to have high values around this temperature due to the large standard deviation in panel (a). In particular for the power curve in panel (d), we see that the power follows the upper bound closely around the center temperature; this is due to the small difference between the two filter heights, see Table~\ref{tab:average_values}. Both the probability distributions for the two outputs, panels (c) and (e), resemble normal distributions. In Table~\ref{tab:average_values}, it is clear that the feedback gives higher outputs but the values are fairly close to the no-feedback curves optimized for $\Tabs = 20\Tcol$---hardly surprising given that the normal distribution is defined around $\Tabs = 20 \Tcol$. 

Lastly, Fig.~\ref{fig:uniform_results} shows results where the feedback mechanism is optimized for a uniform temperature distribution. Since every temperature is equally likely, the optimization sets the second filter height higher than for the other distributions, which lets the feedback curves saturate the upper bounds for temperatures near the end of the range, where the achievable power and potential are the highest. However, it is also beneficial to have a large difference between the filter heights, as apparent in Table~\ref{tab:average_values}, since setting the first height too high would overly suppress quantities at low temperatures. The probability distributions in panel (c) and (e) are now just the distributions of the optimized quantities over temperature. The spread is larger in comparison to the earlier distributions, meaning that the variance is higher. In particular, the largest peaks are near zero for both quantities, meaning that near-zero outputs are relatively likely compared to useful outputs.

\begin{table}[tbh]
    \centering
    \caption{Average powers and potentials at $\tcharge$ for the three temperature distributions, see Fig.~\ref{fig:add_results}, over the temperature range $\Tabs \in [1.5, 30] \Tcol$.  Subscripts $10\Tcol$ and $20\Tcol$ correspond to no-feedback devices where the filter height is set to maximize the quantity at the indicated temperature, before averaging. Parameters are found in Fig.~\ref{paper-fig:ideal_result} in the main paper. Subscript $\mathrm{opt}$ corresponds to devices with feedback where the parameters are optimized to maximize the average quantity. These parameters are also listed. Note that the average quantities for the optimized feedback curves are all larger than values for nonfeedback ones for the same distribution. Powers in unit $(\kB\Tcol)^2/\hbar$, potentials and parameters in unit $(\kB\Tcol)$.}
    \begin{tabular}{|l|cccccc|cccccc|} \hline
    & & & & \multicolumn{3}{|c|}{Parameters $\bar P_\mathrm{opt}$} & & & & \multicolumn{3}{|c|}{Parameters $\overline{(\mu|_{t_\mathrm{max}})}_{\mathrm{opt}}$} \\ \hline
         Distribution & $\bar P_{10 \Tcol}$ & $\bar P_{20 \Tcol}$ & $\bar P_\mathrm{opt}$ &$\enone$ &$\entwo$ &$\edot$ &$\overline{(\mu|_{t_\mathrm{max}})}_{10 T_\mathrm{c}}$ &  $\overline{(\mu|_{t_\mathrm{max}})}_{20 T_\mathrm{c}}$ & $\overline{(\mu|_{t_\mathrm{max}})}_{\mathrm{opt}}$&$\enone$ &$\entwo$ &$\edot$\\ \hline 
         Bimodal & 7.51 & 8.62 & 9.53 &6.68 &13.10&  8.73 & 18.61 & 17.79 & 24.90 & 13.95 & 33.67&  17.05\\
         Normal & 11.60 & 15.44 & 15.67 &12.21 &14.91 &13.10& 22.32 & 29.23 & 33.74 & 19.10 &35.91 &21.26\\
         Uniform  &8.77 & 10.85 & 11.65 & 8.41 &16.05 & 11.13&17.27 & 18.76 & 23.98 &14.45 & 39.89&18.22\\ \hline 
    \end{tabular}
    \label{tab:average_values}
\end{table}

\bibliography{refs}